\documentclass[sigconf,10pt,nonacm]{acmart}
\usepackage{graphicx}
\usepackage{epstopdf}
\usepackage{float}
\usepackage{balance}
\usepackage{comment}
\usepackage{amsmath}
\usepackage{multirow}
\usepackage{amsmath}
\usepackage{xspace}
\usepackage{xcolor, color, soul}
\usepackage{url}
\usepackage{pifont}
\usepackage{wasysym}
\usepackage{tablefootnote}
\usepackage[flushleft]{threeparttable}
\usepackage{footnote}
\makesavenoteenv{tabular}
\DeclareMathOperator*{\argmin}{argmin}

\usepackage{breakurl}

\newcommand{\new}{\vskip 0.75em plus 0.15em minus .1em}

\newcommand\todo[1]{\textcolor{black}{#1}}

\newcommand\todoyang[1]{\textcolor{black}{#1}}
\newcommand\todoniru[1]{\textcolor{black}{#1}}
\newcommand\commentHarsh[1]{\textcolor{black}{#1}}
\newcommand{\name}{\textit{WhisperWand}\xspace} 
\newcommand{\squishlist}{
	\begin{list}{$\bullet$}
		{  \setlength{\leftmargin}{+0.05in}
		} }

		\newcommand{\squishend}{
	\end{list} }

\usepackage{soul,color}

\usepackage{mathtools}

\hyphenchar\font=-1
\makeatother

\begin{document}
\title{WhisperWand: Simultaneous Voice and Gesture Tracking Interface}


\author{Yang Bai}
\affiliation{
  \institution{{yangbai8}@umd.edu}
  \institution{University of Maryland College Park}
  \country{}
}

\author{Irtaza Shahid}
\affiliation{
  \institution{irtaza@umd.edu}
  \institution{University of Maryland College Park}
    \country{}
}

\author{Harshvardhan Takawale}
\affiliation{
  \institution{htakawal@umd.edu}
  \institution{University of Maryland College Park}
    \country{}
}

\author{Nirupam Roy}
\affiliation{
    \institution{niruroy@umd.edu}
    \institution{University of Maryland College Park}
      \country{}
}


\begin{abstract}

This paper presents the design and implementation of \name, a comprehensive voice and \todo{motion tracking} interface for voice assistants.
Distinct from prior works, \name is a \todo{precise tracking} interface that \todo{can co-exist with the voice interface on low sampling rate voice assistants}. \todo{Taking handwriting as a specific application,} it can also capture natural strokes and the individualized style of writing \todo{while occupying only a single frequency}. The core technique includes an accurate acoustic ranging method called Cross Frequency Continuous Wave (CFCW) sonar, enabling voice assistants to use ultrasound as a ranging signal while using the regular microphone system of voice assistants as a receiver. 
\todo{We also design a new optimization algorithm that only requires a single frequency for time difference of arrival.}
\name prototype achieves 73 \commentHarsh{$\mu$m} of median error for 1D ranging and 1.4 \commentHarsh{mm} of median error in 3D tracking of an acoustic beacon using the microphone array used in voice assistants. Our implementation of an in-air handwriting interface achieves 94.1\% accuracy with automatic handwriting-to-text software, similar to writing on paper (96.6\%). 
\todo{At the same time, the error rate of voice-based user authentication only increases from $6.26\%$ to $8.28\%$.}
%
%
\end{abstract}
\maketitle




\section{Introduction}
Voice assistants (also commonly known as smart speakers) have become a common household device.
Nearly half (48\%) of US internet users own one or more smart speakers \cite{vastatone} and global sales of such devices are likely to surpass 200 million annually \cite{vastattwo}.
This proliferation gives rise to an ecosystem of innovative acoustic sensing and perception capabilities around these devices beyond traditional speech recognition.
Researchers have developed a spectrum of applications on voice assistants including health monitoring \cite{nandakumar2015contactless, wang2021using,zhang2020your}, gesture recognition \cite{gupta2012soundwave, ruan2016audiogest}, acoustic analytics \cite{ahmed2020preech, vaidya2019you}, and environmental monitoring \cite{voloc, wang2021mavl}.
\new

\todo{
Simultaneous motion tracking on these voice assistants can open up a wide range of new applications.
Precise motion tracking can be used as a handwriting interface for nonvocal communication, two-factor authentication using voice and signature, hand gesture can enable nonvocal interactions to these ubiquitous devices for voice-impaired persons and elderly people. 
Moreover, a simultaneous voice and gesture interface can be an authentication channel using a library of pre-defined hand motions.
However, existing acoustic motion tracking studies cannot be applied on voice assistants that can only support a limited audible bandwidth(e.g., Amazon Echo (0-8kHz)~\cite{samplingrate,shen2020voice,chan2019contactless,taori2019targeted}, Google Nest Audio(0-10kHz)~\cite{googlenestaudio}, Google Nest Mini (0-15.3kHz)~\cite{googlenestmini}, Sonos Move (0-18.5kHz)~\cite{sonosmove}) due to two reasons: (1) To achieve inaudibility, existing studies use the near-ultrasound band (18-24kHz) for motion tracking, which is not supported by low-end voice assistants. (2) Directly applying wide-band tracking on low-end devices interferes with the voice interface.
In this paper, we push the limit of acoustic motion tracking accuracy to a few tens of micro-meter using a single frequency, without any interference to the voice interface. 
}
\new

\begin{figure}[t]
\begin{center}
\fbox{\includegraphics[width=3.15in, height=1.6in]{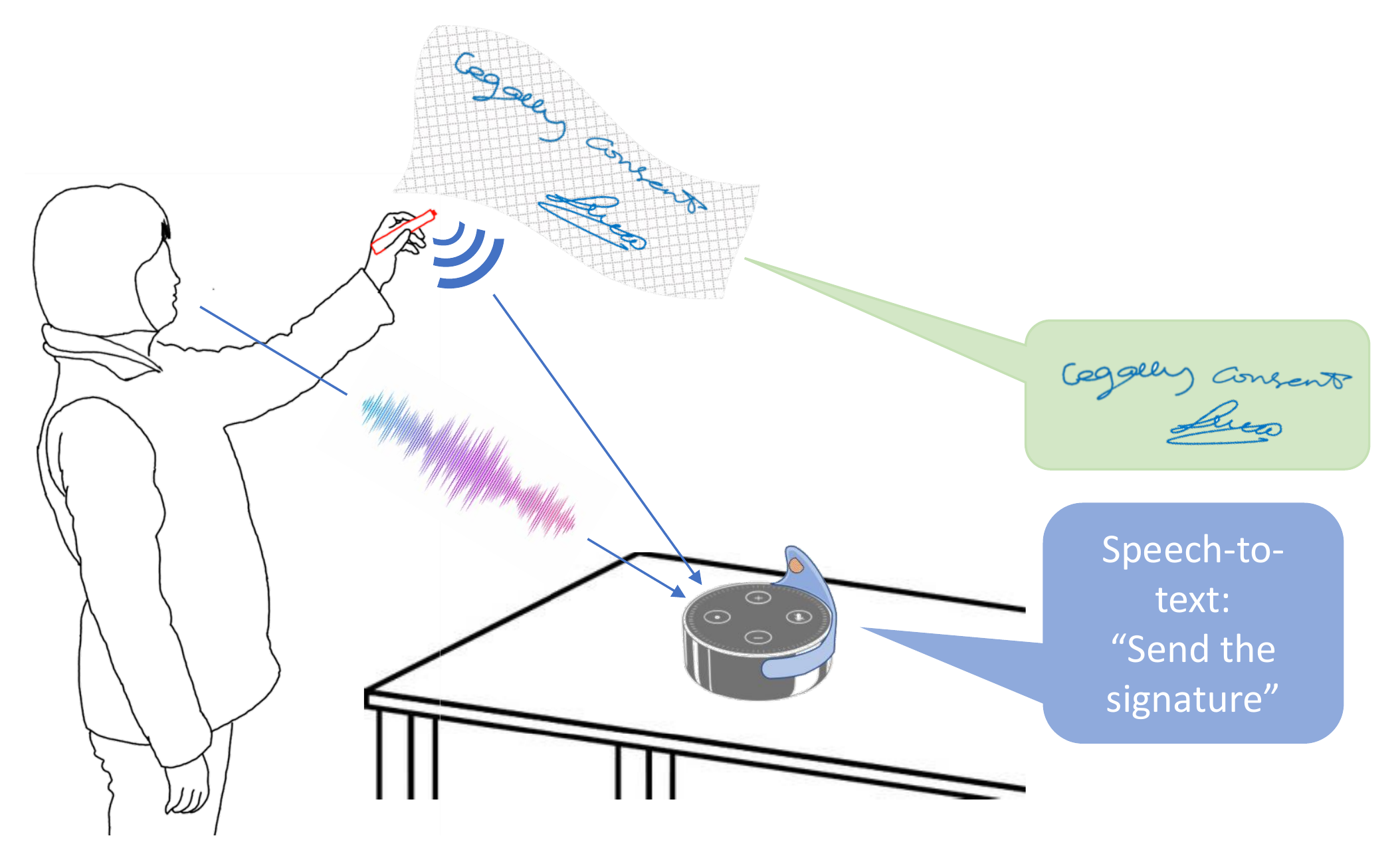}}
\vspace{-0.1in}
\caption{\name enables voice assistants to \todo{capture human voice and track motions simultaneously on a voice assistant that can only support 0-8kHz frequency band. The motion tracking only occupies a single frequency, without interference to the human voice.}}
\vspace{-0.2in}
\label{fig:FirstPageFigure}
\end{center}
\end{figure}

With the advent of smart devices, several techniques have been proposed to enable a virtual \todo{motion tracking} interface in the air \cite{nandakumar2016fingerio, mao2016cat,zhang2017soundtrak,wang2019millisonic}.
\todo{While a lot of applications are possible with these techniques, they are far from being precise enough from supporting applications such as natural writing interface.}
\todo{Existing works also require a wide-band signal for motion tracking, hinder the co-existence with voice interface on low sampling rate devices.
We ask the question, {\em How precisely can we track motion using a small bandwidth or just one frequency?}
}
\new

A popular approach for contactless motion detection repurposes the principles of active sonar.
The Frequency Modulated Continuous Wave (FMCW) sonar technique is particularly effective for detecting small motions due to its high sensitivity and range resolution.
Here an acoustic speaker generates a chirp signal with continuously varying frequency within a fixed bandwidth and the receiving microphone detects the distance of the source from the phase delay of the received chirp.
Unfortunately, the resolution of the FMCW sonar is limited by its bandwidth.
Even if a system does not mind using audible sounds, it can use a maximum bandwidth of around $8$kHz on most of the commodity devices.
Moreover, using the entire audible bandwidth prevents the system from simultaneously processing human voice signal.
\new


This paper presents \name, an inaudible motion tracking interface for voice assistants.
The system shows fine-grained motion detection in a real-world scenario using only a single frequency without affecting the device's voice recognition capabilities.
\name achieves these targets by using a very high-frequency ultrasound signal for ranging.
Although regular voice assistants are not capable of receiving ultrasound signals, we show that the inherent non-linearity present in the off-the-shelf microphone systems provide a natural multiplication operation in the acoustic signal path.
If signals are carefully designed, they can leverage this implicit multiplication operation to shift down a high-frequency signal to the low-frequency band, which the microphone can readily record without any modification to the hardware.
We build on this intuition to develop our {\em Cross-Frequency Continuous Wave (CFCW)} sonar that generates ultrasound for ranging but receives and processes the signal at the regular audible range in the off-the-shelf voice assistants.
The high-frequency signals offer a clear advantage in the resolution of distance tracking while inaudible to human of all ages.
\todo{Moreover, through careful design, CFCW sonar only occupies one frequency at the audible band, minimizing interference the to voice interface.}
Next generations of voice enabled devices are coming with ultrasound speaker for signal generation.
However, for devices with the built-in speakers that cannot produce these high-frequency ultrasound signals, we design an external add-on speaker module capable of producing the required ultrasound signals.
This speaker module operates as a stand-alone external speaker connected through the audio port of the smart speaker.
\new

In addition to the low-latency high-resolution 3D trajectory tracking, 
\todo{\name takes \todo{one of the most precise motions} -- natural strokes of spontaneous handwriting as an application.}
The personalized style of handwriting includes subtle sub-mm movements of the pen tip continuously moving in the air.
Moreover, handwriting recovery will need to translate the arbitrary 3D trajectory of the pen to a 2D writing recognizable by humans and machines.
We address the first challenge in our system, called \name, by developing a novel motion-tracking technique and the second challenge using a sequence of post-processing methods customized for the handwriting interface.
\name applies a geometric mesh parameterization technique to recover the handwriting from an arbitrary virtual 3D surface to a flattened projection on a 2D plane.
The post-processing eliminates any stray trajectory due to pen-lift and unwanted movements of the pen.
\new

\todoyang{This paper focuses on developing a precise in-air \todo{motion tracking} system \todo{for voice assistants that can co-exist with the voice interface.}
We assess the limits of ranging and localization with CFCW sonar 
and application-specific challenges such as removing pen lifts and flattening writing surfaces.}
We made the following contributions in this project:
\squishlist
    \item \todoyang{Designed, implemented, and evaluated a new ranging technique, cross-frequency continuous radar, that can capture minute \todo{motions such as} individualized writing styles in signature. The accuracy of cross-frequency continuous radar is $10\times$ higher than the most precise acoustic-based ranging technique.}
    \item \todo{Developed an inaudible pure tone-based motion tracking system that can co-exist with voice interface on low sampling rate voice assistants that can only support a frequency band below 8kHz.}
    \item Designed a series of processing methods to recover handwriting from a 3D trajectory drawn in the air. The output produces handwriting that retains the individualized writing styles and is legible to humans as well as machines.
    \item Developed a hardware/software prototype of the end-to-end system. We plan to make the designs and codes open-source for the community to reproduce, evaluate, and build on the \name system.
\squishend

\section{Application scenarios}
{\bf Non-verbal interaction interface:}
\todo{
Co-exist with voice interface, \name can be a means for nonverbal communication with voice interfaces for interacting inconspicuously or even covertly.
The user can write comments or draw symbols while listening to a particular portion of a song and this note will be saved on the timeline.
The stylus for writing in the 3D space can be a controller for complex interactions with a voice assistant useful for gaming or gesture-based password.
It can enable a writing interface for taking quick handwritten notes on voice assistants. One example is annotating a live audio (e.g., a song or a speech) played by the speaker.
}
\new
{\bf Authentication with physical signature:}
\todo{
It is natural to wonder {\em if a voice assistant can capture a reliable physical signature, will it open electronic-signature services on this new platform?}
Today voice assistants can read out emails to us, make online payments, and even read bedtime stories.
We can imagine in the future a voice assistant reading out a legal document to a user and collecting a signature for acknowledgment or on the receipt of a transaction.
This sign-in-the-air interface can also verify liveness and enable multi-factor authentication for a secure session of interaction with the voice assistant.
}
\new

\textbf{Potential applications with CFCW sonar:} In addition to \todo{interaction  and handwriting interface}, the core ranging and localization with the cross-frequency sonar can lead to applications that require high-resolution tracking.
Examples include tele-operation and remote surgery with precise hand motion tracking, real-time orientation estimations for AR/VR systems, and bodily vibration and tremor detection.
We look forward to applying CFCW to more applications requiring precise motion tracking.


\section{Primer: Location from phase}


Distance estimation is a fundamental building block of localization, and its accuracy depends on the Time of Flight (ToF) estimation.
The phase of a coherent signal is a measurable parameter that continuously changes over time.
Therefore, if the phase of the received signal is estimated correctly, it can tell the time delay of signal propagation or the ToF from its origin at the beacon. 
When transmitting a pure tone, the received signal can be represented as $S_R = sin(2\pi f(t-t_p))$.
Here $f$ is the frequency of the signal, and $t_p$ is the time delay of propagation.
Through analyzing the phase of recorded signal $2 \pi f t_p$, we can calculate the time of arrival $t_p$.
The phase shift of two adjacent samples can be formulated as $\Delta \phi = 2\pi f \Delta d/c$, where $\Delta d$ is the distance change within the time period and c is the speed of sound.


\subsection{Advantage of pure-tone based ranging}
\todo{
FMCW is a widely accepted technique for ranging and localization. 
However, it has two limitations.
FMCW transmits a chirp that occupies a wide frequency band, and the resolution of distance tracking is proportional to $\frac{c}{B}$, where $c$ is speed of sound and $B$ is the bandwidth of chirp.
To achieve higher resolution, a wide frequency band is occupied. \commentHarsh{ If the band overlaps with the 0-8kHz frequency band of voice, it disables the voice interface.}
Existing works use near-ultrasound frequency band for the chirp, but not all the devices support near-ultrasound frequency band. Voice assistants like Amazon Alexa and Amazon Echo only support 0-8kHz frequency band.
Similar to FMCW, Time-of-Arrival (ToA) and Time-Difference-of-Arrival (TDoA) also require a wide frequency band for precise correlation.
The advantage of pure-tone based ranging is that it only requires one frequency, minimizing the interference with voice interface and other applications requiring a wide frequency band.
}

\begin{figure}[t]
\begin{center}
\includegraphics[width=3.4in]{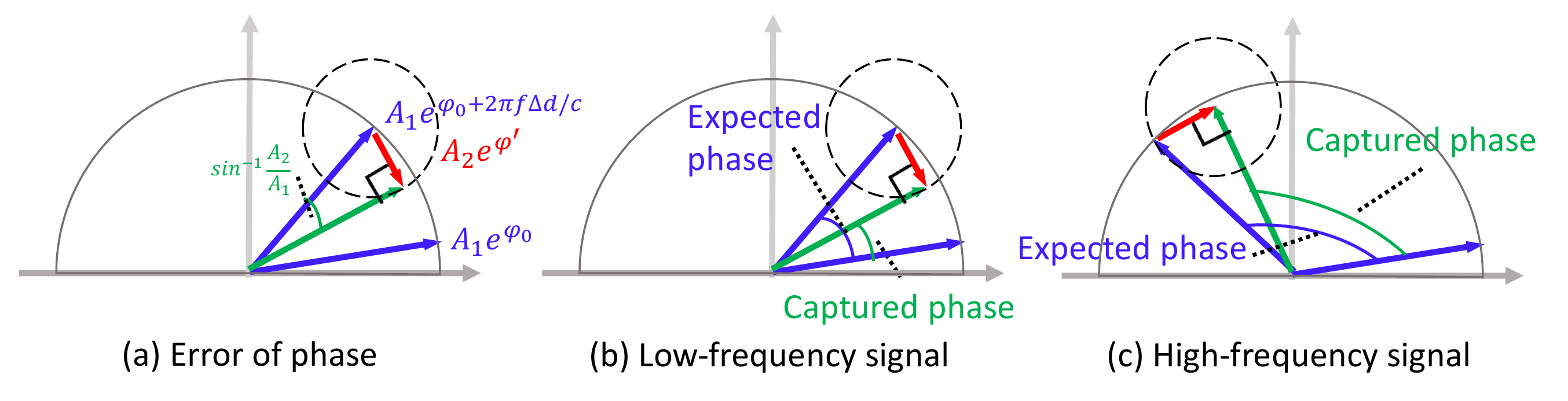}
\vspace{-0.2in}
\caption{{\todo{The phase offset caused by interference leads to smaller distance offset on high-frequency signal over low-frequency signal.}}}
\vspace{-0.2in}
\label{fig:phase_offset}
\end{center}
\end{figure}

\subsection{Advantages of high-frequency signal}
\todo{
{\bf Distance resolution is proportional to the frequency.}
The sensitivity of distance measurement using phase depends on the wavelength of the signal as the phase changes $2\pi$ radian per wavelength.
With same distance change, a larger phase change is shown in higher frequency signal.
The microphones of the voice assistants operate at the low-frequency audible signals of longer wavelength leading to poor ranging and localization performance. 
\new
{\bf High-frequency signal leads to smaller distance offset under interference.} Let us assume that the interference caused by environmental noise or multipath creates an indirect path with a lower amplitude than the direct path in phase domain. 
With same distance change, Figure~\ref{fig:phase_offset} shows the complex representation of phase change for both high-frequency and low-frequency signal. The line-of-sight path is shown in blue and the interference is shown in red.
The sum of the signal in green is the captured phase at the receiver side.
Suppose the object moves in distance $d$, the theoretical phase shift is $e^{2\pi fd/c}$.
The maximum phase error $\Delta \phi$ caused by the interference is $sin^{-1} (\frac{A_2}{A_1})$~\cite{wang2019millisonic}.
If we convert the phase offset to distance offset, $\Delta d = \frac{\Delta \phi c}{2 \pi f} $.
As shown in Figure~\ref{fig:phase_offset} (b)(c), the key observation is the distance offset caused by interference is proportional to frequency.
In other words, high-frequency signal is more robust to interference in distance tracking.
Therefore, we use high frequency signal for ranging.
}

\subsection{\todoyang{Challenges of pure tone-based ranging}}

{\bf Vulnerable to multipath distortion.}
Unmodulated pure tone-based systems suffer from an environmental multipath effect that degrades the phase accuracy of the signal.
\todoyang{Although pure tone ranging is accurate, multipath fading severely impacts the accuracy. The non-line-of-sight signals fall in the same frequency. This inseparable overlap interferes with the phase values, leading to inaccurate distance estimation.} 
\new
\commentHarsh{{\bf Inability to obtain initial distance.}} \todo{While pure-tone based ranging can capture precise distance change, it cannot capture the absolute initial distance due to phase wrapping. The phase of absolute distance is $2n\pi + \phi$, phase tracking can only capture $\phi$, without information of the number of wrapping $n$.}
We address these limitations in our novel ranging technique explained next. 

\section{Cross-frequency sonar design}
\name develops an acoustic ranging method that uses two completely separate bands for the transmit signal and received signal -- leading to the ranging technique we call Cross-Frequency Continuous Wave (CFCW) sonar.
If works, CFCW sonar can introduce several crucial advantages to sensing applications with these household devices.
{\bf (a) (Accuracy)} Given it can leverage high-frequency ultrasound signals for ranging, it can have orders of magnitude higher accuracy compared to audible frequency-based techniques.
{\bf (b) (Inaudibility)} It allows accessing a wide band of inaudible frequencies with regular microphones, while existing techniques can only use a small band of `near-ultrasound' frequencies between 18-24 kHz.
{\bf (c) (Power and hardware simplicity)} \todo{CFCW sonar operates at a low frequency which requires lower sampling rates. It offers similar accuracy to ultrasound with low complexity hardware as the power consumption and processing and storage requirements of the platform increase with the frequency of operation.} 
\new

The CFCW sonar leverages implicit frequency translation possible in regular microphones.
It carefully designs its ultra-sound transmit signals such that they can leverage the fundamental nonlinearity in microphones for automatic down conversion of the signals to the low-frequency recording range of the microphones.
While nonlinear frequency conversion enables the core CFCW technique, it requires further signal design to eliminate environmental effects for effective distance \todo{tracking and start point detection} for \name.
We explain the step-by-step signal model next.
\new

\subsection{\todo{Implicit frequency and phase translation}}
\begin{figure}[t]
\begin{center}
\includegraphics[width=2.8in]{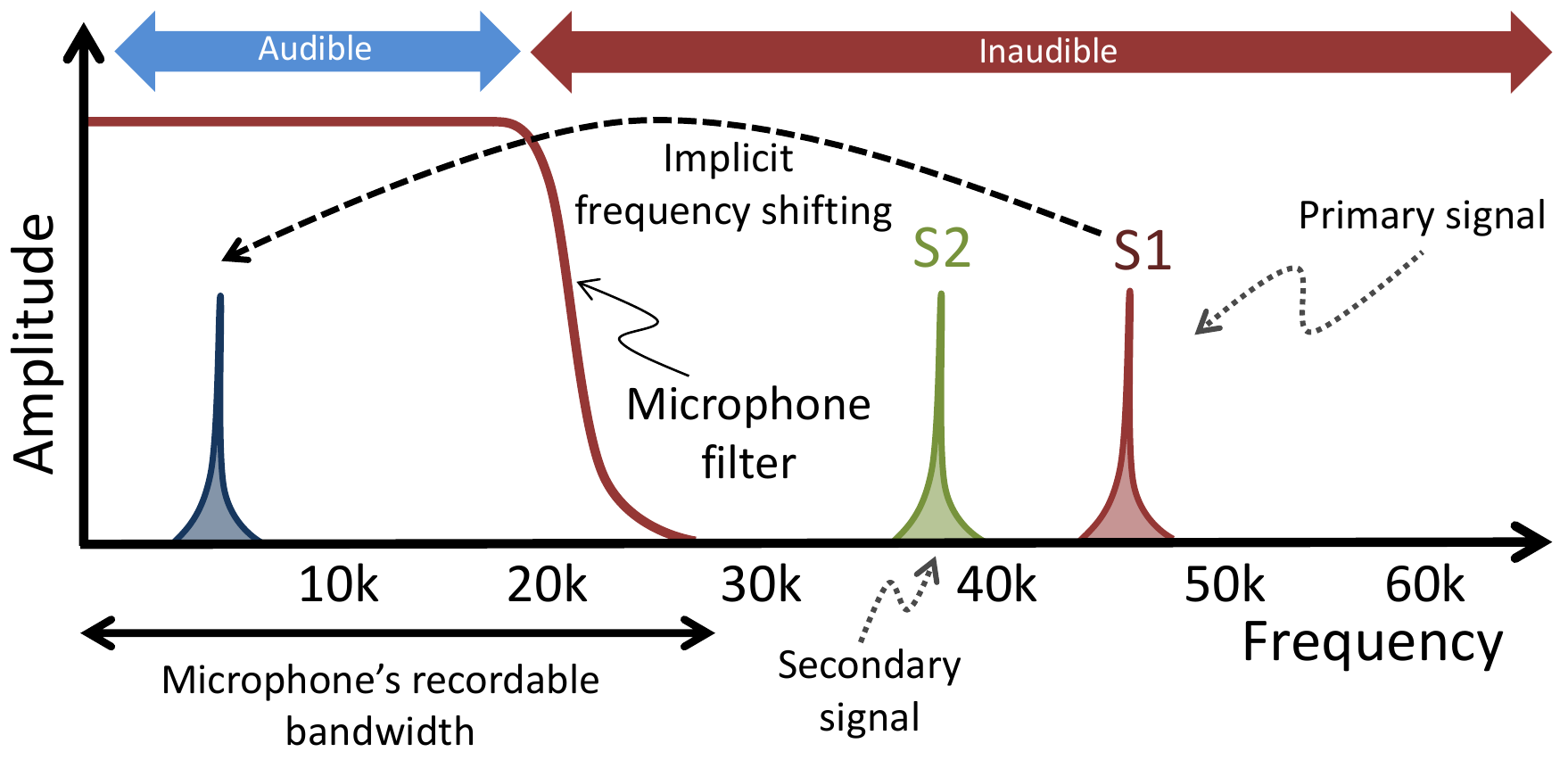}
\vspace{-0.1in}
\caption{\todoyang{The non-linearity of the microphone multiplies ultrasound signals implicitly shifts the frequency to the audible band.}}
\vspace{-0.1in}
\label{fig:nonlinearity}
\end{center}
\end{figure}

\begin{figure}[h]
\begin{center}
\includegraphics[width=1.5in]{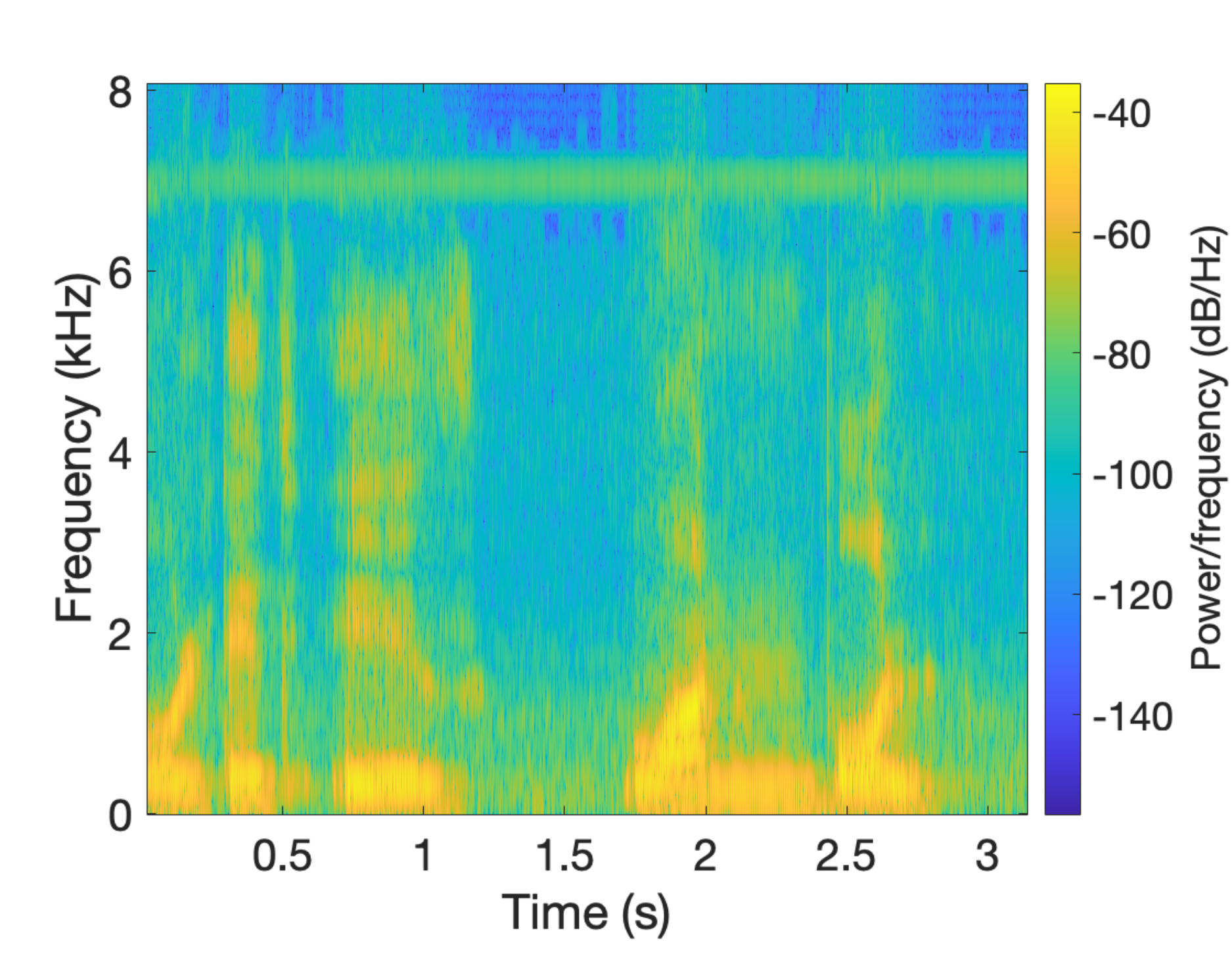}
\includegraphics[width=1.5in]{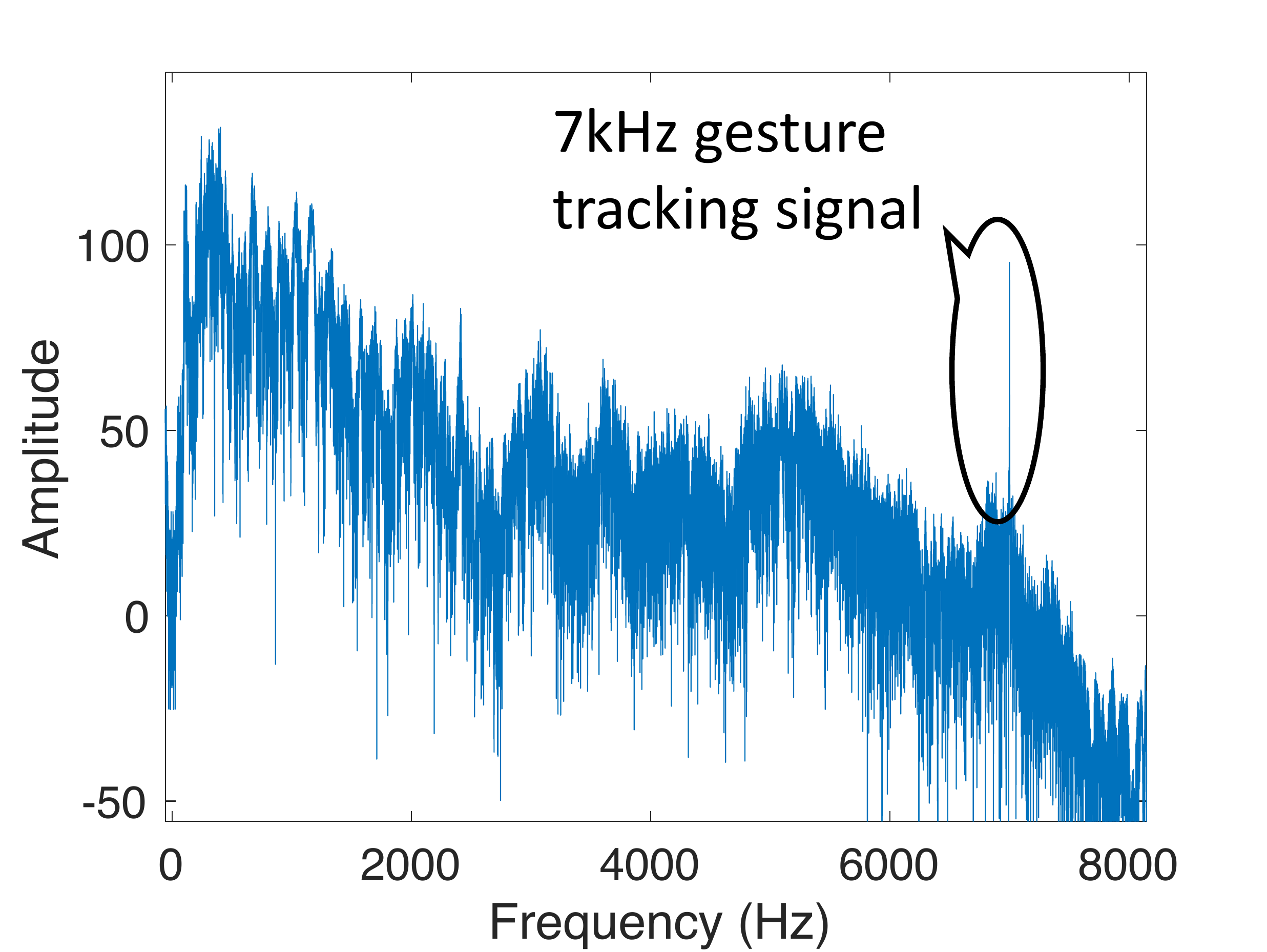}
\vspace{-0.1in}
\caption{(a) Spectrogram and (b) FFT of the signal captured by Alexa. \todo{Motion tracking only occupies 7kHz frequency.}}
\vspace{-0.2in}
\label{fig:spectrogram_alexa}
\end{center}
\end{figure}

{\bf Microphone non-linearity primer: }
Commonly microphone is a linear system, which means the signal recorded by the microphone is a linear combination of the input signals. If the input signal is $S$, the recorded signal $S_{out}$ is $S_{out} = A_1S$,
where $A_1$ is the complex gain. 
However, as introduced in Backdoor~\cite{roy2017backdoor}, the microphone shows non-linearity when the frequency of the input signal is above $25kHz$.
The recorded signal $S_{out}$ can be modeled as $S_{out} = \sum_{n=1}^{N} A_n S^n$.
Since the third and upper terms are too weak and thus can be ignored, the $S_{out}$ can be represented as $S_{out} = A_1 S + A_2 S^2$.
As shown in Figure~\ref{fig:nonlinearity}, we transmit two ultrasound pure tones together as $S = sin(2\pi f_1 t) + sin(2\pi f_2 t)$, where $f_1$ and $f_2$ are 45kHz and 38kHz.
With nonlinearity of microphone, the recorded signal also includes the components produced by $S^2$, which are $f_1 + f_2$ and $f_1 - f_2$. While $f_1 + f_2$ is higher than the threshold that can be recorded by the off-the-shelf microphones, $f_1 - f_2$ can be recorded with proper selection of frequencies, as shown in Figure \ref{fig:nonlinearity}.
\new
\todo{
Interestingly, if phase change is happened in frequency $f_1$, the phase change can be mapped to the down-converted frequency $f_1 - f_2$. 
As $S = sin(2\pi f_1 t + \phi) + sin(2\pi f_2 t)$, the received signal at frequency $f_1 - f_2$ is $sin(2\pi (f_1 - f_2)t + \phi)$.
Therefore, we map the phase of frequency $f_1$ on $f_1 - f_2$ by down-converting the frequency using non-linearity of microphone.
Here we call $f_1$ as primary channel and $f_2$ as secondary channel.}
The primary signal source act as an acoustic beacon attached to the target object.
The secondary signal source is fixed using an add-on module to the voice assistant.
The secondary source emits a low-power signal that illuminates only the microphone array of the voice assistant.
These signals combine inside the microphones to create a different signal that carries the phase offset between the primary and secondary tones.
The phase difference is due to the distance of the target from the voice assistant.
We verify if an off-the-shelf voice assistant can show such implicit frequency shift. We use an Amazon Echo Dot device to simultaneously record the human voice and the shifted tracking signal successfully as shown in Figure~\ref{fig:spectrogram_alexa}.

\subsection{Multipath avoidance}

\name relies on the accurate phase estimation of the signal for ToF measurement, however phase of the received signal can be severely affected by environmental effects of which multipath signal propagation is the most significant.
Multipath is a natural phenomenon where a signal, after leaving the transmitter, reflects off objects in the environment to create replicas and the replicas propagate through paths of different delays before combining at the receiver.
The lengths of these individual paths decide the phase delays of the replicas and therefore their superimposition leads to an unknown amplitude and phase of the received signal.
For location tracking, \name requires the phase of the direct line-of-sight (LOS) path, but any non-line-of-sight (NLOS) path can introduce an unpredictable error to the phase.
However, note that there is only one LOS path, and its length is shorter than all NLOS paths.
NLOS paths are indirect paths and therefore arrive at the receiver later than the LOS signal.
This provides a window of opportunity to extract the undistorted phase from the LOS path before the multipath replicas superimpose on this signal as explained next.
Potentially the phase estimation system can use a part of the signal clear from multipath superposition to get an accurate phase of the transmit signal.
\new


Theoretically, in a pulse-based probing signal, there exists a time window when only the LOS signal is present at the receiving sensor.
The length of this time window WIN-LOS
is equal to the delay of the first NLOS path which is an unknown environmental parameter.
The smallest delay of the NLOS path, which leads to the smallest size of the time window, is related to the distance of the closest large objects to the device.
Another parameter that impacts the worst-case estimate of the time-window WIN-LOS is the FFT resolution.
\name estimates the phase by first calculating the FFT coefficients of the time series data of length WIN-LOS seconds.
This signal is sampled by the microphone at 16 kHz (standard sample rate of \todoniru{Alexa}) after the nonlinear conversion of the ultrasound signal.
A too-short WIN-LOS will lead to wider FFT bins combing a wider band of frequencies to the same bin.
At the baseband, the microphone also records ambient sounds, such as human voices and other household noise.
The energy of these sounds is mostly limited within 4 kHz \cite{wang2021mavl} and the CFCW signal is mapped to 7 kHz at the baseband.
Therefore, we limit the FFT bin width to a maximum of 2 kHz, which in turn limits the minimum size of WIN-LOS to 0.5 ms around 17cm in distance.
A pulsed signal can hop to a different frequency within this time to avoid delayed NLOS signals.
As explained next, the CFCW receiver can continuously operate on a constant frequency while the ultrasound-ranging signals hop.
\new




\begin{figure}[t]
\begin{center}
\includegraphics[width=3.1in]{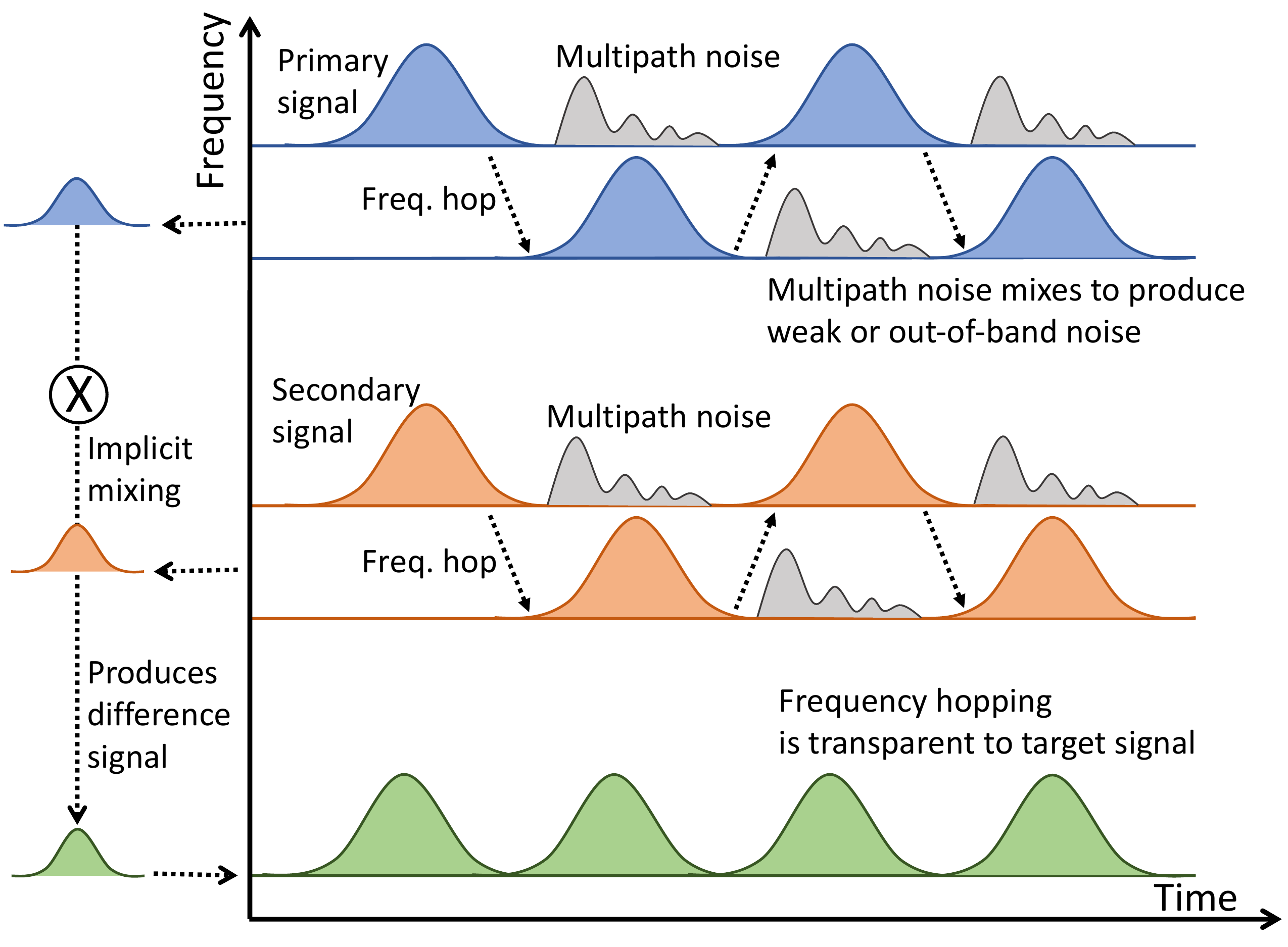}
\vspace{-0.1in}
\caption{\todoniru{An illustration of transparent frequency hopping. Although the transmitted frequencies keep changing, the frequency of the received signal is constant. The spurious signal does not interfere with the frequency of tracking.}}
\vspace{-0.15in}
\label{fig:multipathphaseerror}
\end{center}
\end{figure}

{\bf Transparent frequency hopping. }
In CFCW sonar, the received frequency $f_{rcv}$ is different from the dual frequencies ($f_{sty}$ and $f_{base}$) transmitted in the air and it is equal to the difference between the two transmitted frequencies ($f_{rcv} = f_{sty} - f_{base}$).
Therefore, if the transmitted frequencies change in a synchronized way to maintain a constant difference between them, the received signal will remain constant.
In other words, we can design a frequency hopping for the transmitted signal pair while its effect is transparent to the received signal.
We used this observation to avoid multipath in the CFCW sonar.
\new

The dynamic multipath distortion is primarily caused by the $f_{sty}$ signal source, which is \todo{facing towards the air}, changes location over time, and is of relatively higher power to have stronger NLOS reflections.
The $f_{base}$ signal source, on the other hand, is placed in a static location only a few centimeters away from the microphone array.
Moreover, the power of the $f_{base}$ signal is kept low -- just enough to reach all the microphones in the array -- making NLOS multipath from this signal too weak to cause a significant impact on the received difference signal.
Therefore, if $f_{sty}$ and $f_{base}$ signals hop to new values $\hat{f}_{sty}$ and $\hat{f}_{base}$ respectively, delayed NLOS components of $f_{sty}$ will only mix with the current $\hat{f}_{base}$ to create a spurious signal at frequency ($f_{sty} - \hat{f}_{base}$).
This spurious frequency does not interfere with the received CFCW frequency $f_{rcv}$ leaving it unaffected by the multipath effect.
Figure \ref{fig:multipathphaseerror} shows this transparent frequency hopping-based multipath avoidance where the transmitter's frequencies keep hopping to new values together every 3 ms.
\new

{\bf Phase unwrapping. }\label{sec:phase__unwrapping}
Phase unwrapping is a classical approach to recover the original phase value from the wrapped phase value. 
The phase value is wrapped within $[-\pi, \pi]$, and the goal of phase unwrapping is to reconstruct the continuous phase by removing the "sudden jumps". 
The true phase (i.e., true phase without wrapping) can be reconstructed as long as the difference between the subsequent phases is less than $\pi$.
To satisfy this condition, we need to guarantee the speed is less than $\frac{c}{2f_1 \Delta T}$.
With a $f_{sty}$ 40kHz and $\Delta T$ 3ms, the maximum speed is 1.41m/s.
It is revealed that the peak velocity of hand movement is 2.7m/s, and the peak velocity of hand gesture is 1.8m/s~\cite{degoede2001quickly}. 
Meaning our system can result in an error when the moving speed is high, limiting us to achieve high accuracy with even higher frequency.
To solve this problem, we induce speed to estimate the direction of phase rotation.
Ideally, when the speed was positive in the last samples and was not decreasing dramatically, it is more likely the phase is rotating anti-clockwise. 
Otherwise, when the speed was negative (the direction of movement is opposite) and was not decreasing in the last samples, it is rotating clockwise.
In our system, we only require the absolute difference between subsequent phases is within $2 \pi$, thus the highest speed we can support is 2.83m/s.
To make sure there is an overlap between the two transmitted signals with a supported distance of 70cm, we set a frame rate of 333.
\new
\todoyang{{\bf Why synchronization is not required?}
\todo{We do not require synchronization between transmitters and receivers. 
The reason is even with synchronization, we still cannot find the initial distance by using a single frequency. 
A precise correlation for ToF requires a wide band of frequency.
To get the absolute distance for localization, we design an algorithm to detect the distance at the start point without the requirement of synchronization or a wide band signal as explained next.
}
}

\begin{figure}[t]
\begin{center}
\includegraphics[width=2.6in]{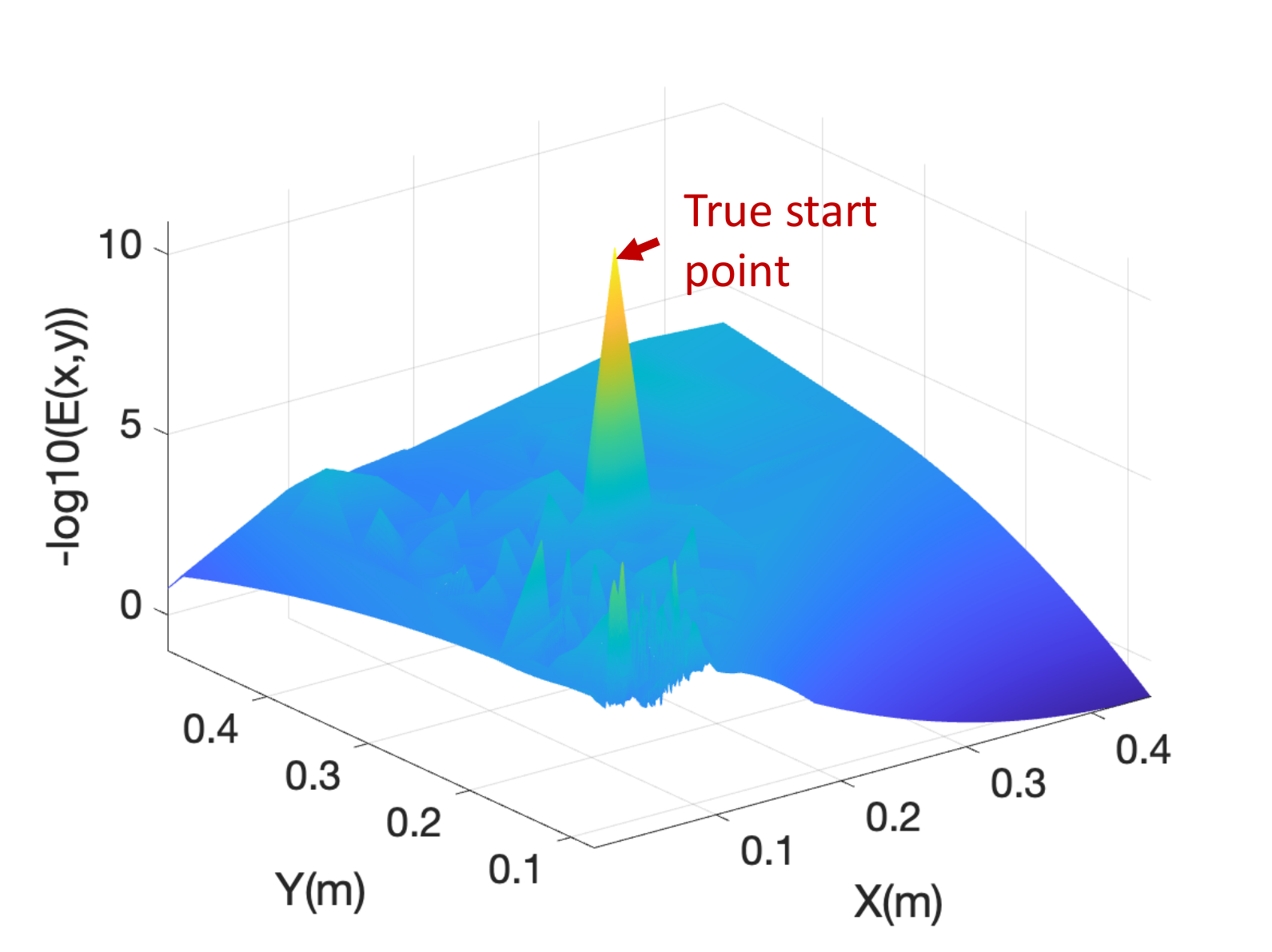}
\vspace{-0.2in}
\caption{Illustration of start point detection technique. Although the optimization function is non-convex, true start point has the global minimum optimization error.}
\vspace{-0.2in}
\label{fig:startpointdetection}
\end{center}
\end{figure}

\new
{\bf Start point detection.} 
\todo{Without synchronization between transmitters and receivers, our start point detection algorithm builds on a traditional approach called Time Difference of Arrival (TDoA).
TDoA approach locates a target at intersections of hyperbolas or hyperboloids that are generated with foci at each fixed receiver of a pair.
Existing algorithm of estimating the time delay between two receivers is to find the time samples that have highest correlation between the transmitted and received signals.
However, the time delay is not precise enough due to two reasons: (1) The time resolution of the correlation is limited by the sampling rate of received signal. (2) A precise correlation requires a sharp pulse as transmit signal, which requires a wide frequency band. Unfortunately, using wide frequency band hints the co-existence between voice interface and tracking for low sampling rate devices like voice assistants.
Since the distance between the microphones is only 3.6cm on voice assistants, the distance drift causes even larger error on location estimation.}
\new

\todo{To address the issue mentioned above, we take advantage of the precise phase from CFCW sonar.
Suppose the wrapped phase difference captured by a microphone pair is $\theta_{ij}$, and the true phase difference is $2n\pi + \theta_{ij}$, where $n$ is the number of wrapping. 
Based on the theory of triangle, the maximum distance difference from speaker to two microphones is the distance between two microphones.
Thus, suppose the range between the pair of microphones is $d$ and the wave length of signal is $\lambda$, we can define the range of $n$ as $n \in [-\frac{d}{\lambda}, \frac{d}{\lambda}]$, where n is an integer.
We find that with multiple pairs of microphones, only when the vector $N$ map with the ground truth, the hyperbolas can cross on the same point, i.e., the start point.
In simulation, we loop through every possible $N$ for microphone pairs to optimize the location $P$ and the corresponding error $E(P)$.
The optimized location $\hat{P}$ is
\begin{equation}
    \hat{P} = \argmin_{P} \sum_{i \neq j} \left\| \lambda (N_{ij} + \frac{\theta_{ij}}{2\pi}) - c\tau_{ij}(P) \right\|^2
\end{equation}
where $i$ and $j$ represents the ids of microphones, $\lambda$ is the wave length, $N_{ij}$ is the number of phase wrapping for time difference of arrival $\tau_{ij}$, $c$ is speed of sound, and $\theta_{ij}$ is the captured phase difference between two microphones, i.e., $\theta_i - \theta_j$. 
As shown in Figure~\ref{fig:startpointdetection}, although there are many other replica peaks, only the true start point has the minimum optimization error.
Therefore, our optimization problem aims to find $\hat{N}$ and $\hat{P}$ that gives the minimum error.
Since bruteforcely looping all possible $N$ is computationally heavy, the optimization boils down to the minimization of 
\begin{equation}
    \hat{P},\hat{N} = \argmin_{P,N} \sum_{i \neq j} \left\| \lambda (N_{ij} + \frac{\theta_{ij}}{2\pi}) - c\tau_{ij}(P) \right\|^2
\end{equation}
where $N \in [\frac{-D}{\lambda}, {\frac{D}{\lambda}}]$ and $D$ is the vector of distance between each pair of microphones. As shown in Figure~\ref{fig:startpointdetection}, this objective function is non-convex, with many local minima. Moreover, this function is discrete, with $N$ as integers. 
To solve this challenge, we apply a non-convex optimization algorithm called genetic algorithm~\cite{mirjalili2019genetic} that also accept integer parameters.
By minimizing the loss, we optimize the $N$ and $P$ simultaneously to find the true start point with global minima.
This approach frees up the requirement of using a wide band signal to capture the time delay for TDoA.
}


\section{Application-specific design}
Precise motion tracking on voice assistants can benefit a wide range of existing applications and open up possibilities for new ones.
Several non-verbal human-machine interfaces can be developed with gesture and motion tracking.
\todo{We have developed natural handwriting detection as the representative application.}
Detection of words written in the air is a classic application explored with a variety of sensing modalities.
We pushed the boundary of this application by not only enabling fine-grained motion tracking required for subtle strokes created during sub-mm movements of the fingers, but also developed techniques for post-processing the data so that even the handwriting can be verified from the notes.
Interestingly, during the process of motion detection, \name's signal does not interfere with the voice signal and therefore the voice assistant can perform its regular operations while simultaneously capturing the writing.
\new

\begin{figure}[hbt]
\begin{center}
\vspace{-0.2in}
\includegraphics[width=1.5in]{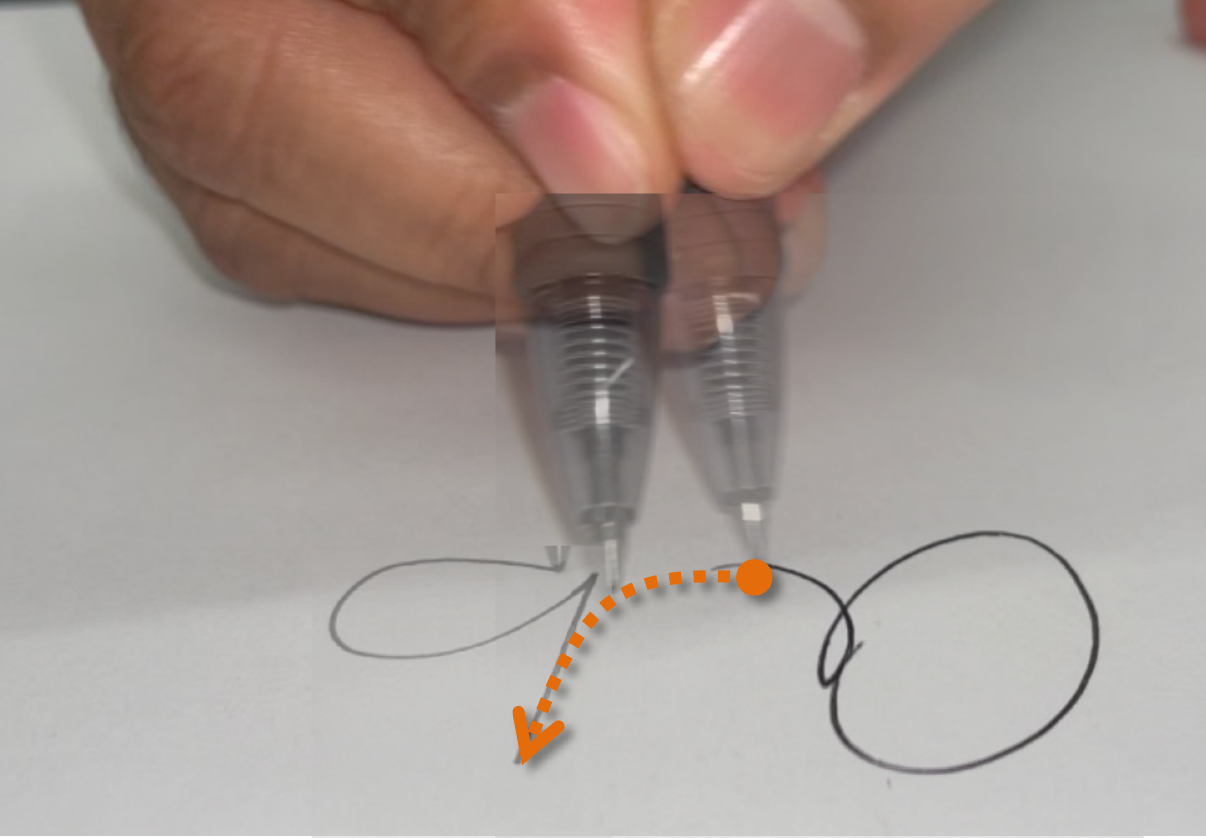}
\includegraphics[width=1.5in]{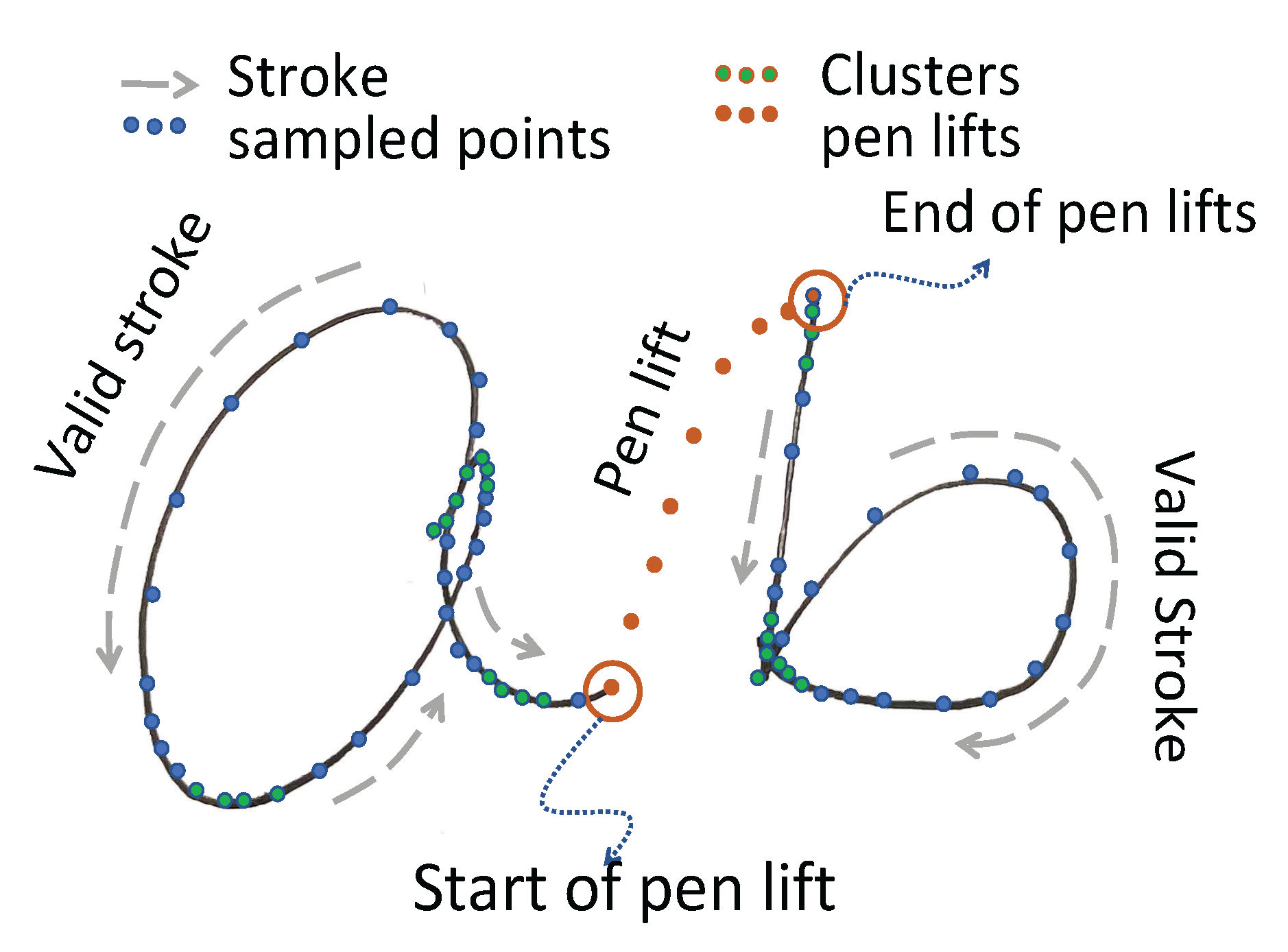}
\caption{\todoniru{An analysis of the handwriting process. Each character has two to four clusters that have slower moving speed than the strokes and pen lifts.}}
\vspace{-0.2in}
\label{fig:cluster_handdraw}
\end{center}
\end{figure}

\name detects the pen strokes by localizing the acoustic source (\todoniru{the primary signal}). 
However, recovering meaningful writing from the continuous trajectory of the pen tip will require addressing two challenges.
{\bf (a) Pen-lift elimination:}  While writing on paper or other solid surfaces, we often lift the tip to jump to the next stroke which is disconnected from the current one, as shown in Figure~\ref{fig:cluster_handdraw}. This event is transparent while writing on a solid surface like a paper or a tablet screen as it does not generate any strokes.
However, 3D tracking-based writing detection cannot naturally distinguish a pen-lift event from the actual strokes. 
{\bf (b) Flattening virtual surface:} Given the user can write at any angle on her assumed surface in the air and the surface is not flat, the projection of the 3D trajectory on the horizontal plane leads to distortion of the strokes. 
The distortion often makes the writing illegible to both humans and \todoniru{automatic text recognizing software}.


\subsection{Localization in 3D space}\label{sec:loc_algo}
Distances from at least three spatially separated microphones enable us to apply the traditional trilateration method \cite{doukhnitch2008efficient} to find the 3D location of the target in space.
We use distance measurements from all the available microphones for location estimation using a multi-lateration technique.
Therefore, the location of the target should be somewhere on a sphere of radius $d_m$, centered at the microphone's location.
Theoretically, we should be able to find the location of the target by solving the system of $N$ equations.
However, in practical systems, the individual distance measurements are not perfect due to noise, and therefore, the spheres do not have a common solution.
We use the following optimization to find the estimated location of the target in 3D space.
It essentially minimizes the distance of the estimated location from all the spheres.
\begin{equation}
\argmin_{x,y,z} \sum_{i=1}^{N} \sqrt {(x - x_i)^2 + (y - y_i)^2 + (z - z_i)^2 - d^2_i}
\end{equation}
\new


\subsection{Removing pen-lift from virtual surface} 
\begin{figure}[h]
\begin{center}
    \includegraphics[width=1.5in]{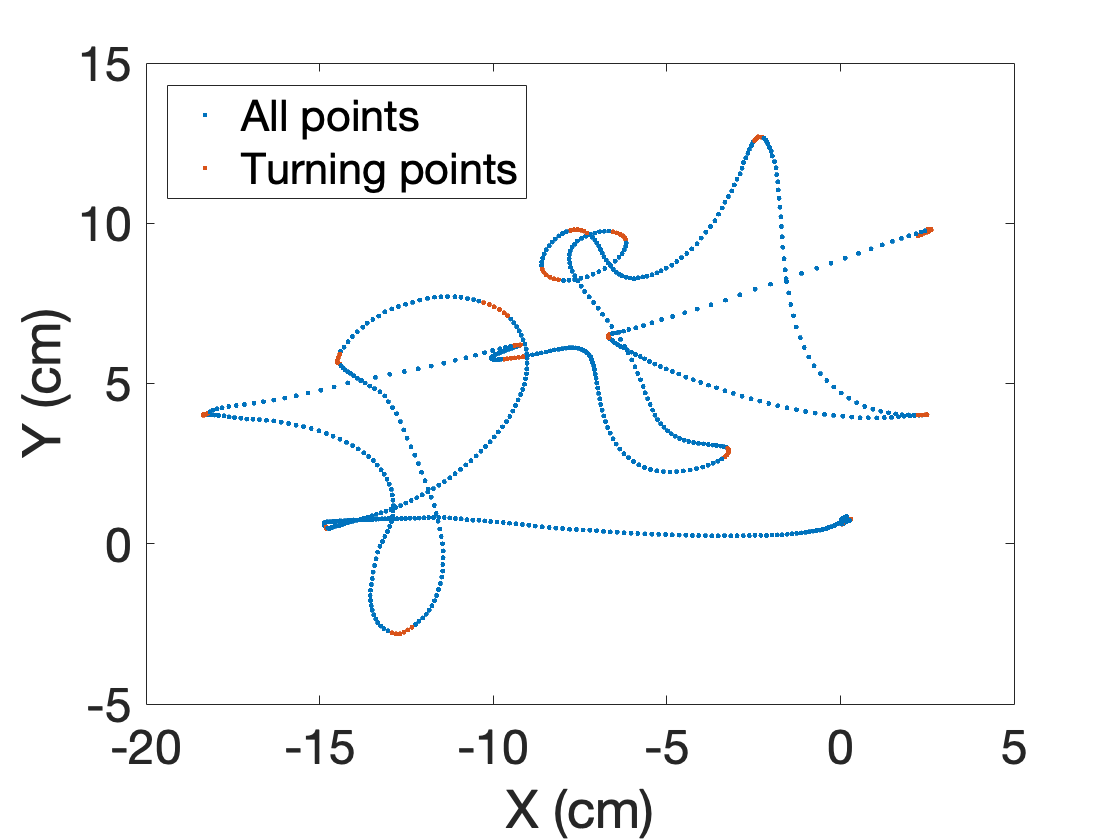}
    \includegraphics[width=1.5in]{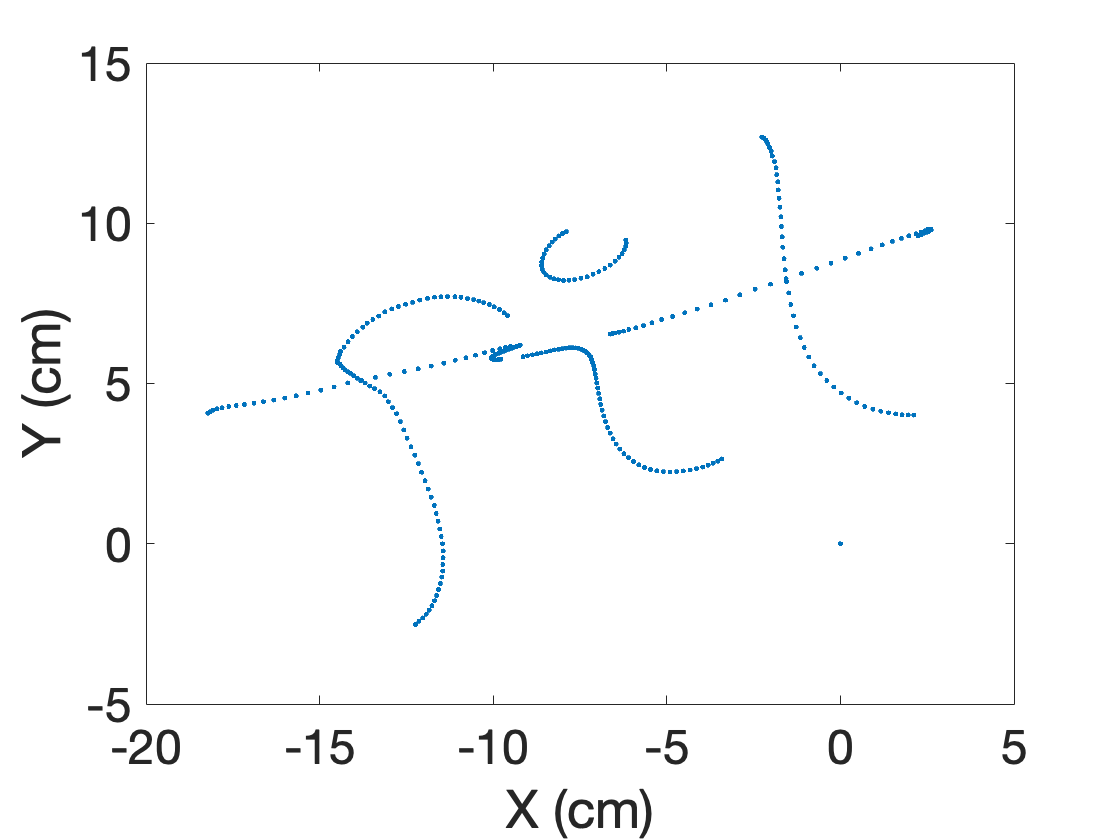}
    \vspace{-0.1in}
    \caption{An example of pen lift removal of the word "fit".}
    \vspace{-0.2in}
    \label{fig:penlift_example}
\end{center}
\end{figure}

The 3D trajectory contains both the actual pen strokes, which are a part of the writing, as well as the stray movements, like pen-lifts.
As shown in Figure \ref{fig:penlift_example}(a), the written word `fit' is unrecognizable from the raw trajectory of the pen because of unwanted marks due to pen-lifts in between two disconnected strokes.
To avoid this error, we will first need to identify the accurate start and end of the pen-lift trajectories.

\new

Handwritten letters and drawings can be viewed as a collection of stroke segments of different curvatures and lengths.
Each of these segments starts and ends with a change in direction of the stroke.
While it is difficult to quantify an angle or sharpness in the change of directions of segments due to various writing styles and user-dependent features, the velocity of the pen movement offers a reliable feature.
The segment-by-segment annotated handwriting reveals that the speed of the pen comes to a low value at the end of a segment and just before the beginning of the next, as shown in Figure \ref{fig:cluster_handdraw}.
We adopted this intuition by first calculating the velocity of the pen between every two samples of location on the trajectory.
\new

A cluster of 3D location points that corresponds to the local minima of the time-varying velocity indicates the start or end of a segment.
Note that these clusters i.e., the ends of the segments are necessarily belong to the imaginary writing plane in the air.
Now it is left to identify a pair of such clusters indicating the start/end of a pen-lift.
To this end, first, we fit a triangular plane to each three of the clusters.
A stroke segment joining two clusters of this triangle is either entirely a valid stroke or a pen-lift.
We calculate the orthogonal distance of all points of a segment from this plane to identify a pen-lift segment by its off-plane movement.
In a few rare cases, the user slows down during a pen-lift movement which erroneously produces a cluster on a pen-lift trajectory.
Our algorithm mistakenly includes these pen-lift trajectories as valid strokes.
We correct this by fitting a third-order plane through all the clusters.
Given the variations in the imaginary writing, a plane is smooth corresponding to sudden movements for pen-lifts, the spurious clusters on the pen-lift trajectory stand out from this fitted plane.
These clusters are then pruned before applying the triangle fitting mentioned above.
\new

\subsection{Flattening writing surface}
\name aims to develop a natural writing interface on the voice assistants and therefore, does not restrict the orientation of the writing plane or the writing style.
The user can simply write in the air near the voice assistant and \name attempts to produce the writing (or drawing) in a human and machine-readable form.
A simple projection of the trajectory \todo{on to} the horizontal plane can severely squeeze and distort the writing.

\new
For the conversion of the 3D trajectory to 2D writing, we first identify the writing surface from the 3D point cloud of the trajectory points by fitting a third-order surface to them.
This smooth surface is the assumed virtual plane on which the user has written.
Next, we take an orthogonal projection of the points from the 3D pen trajectory on this surface.
Note that the pen-lift and other stray marks are already removed from this trajectory data in the previous step.
The next step is to project this arbitrary writing surface, along with the writings, on the 2D horizontal plane.
The curves on the writing surface will require careful flattening to avoid any distortion of the written strokes during projection.
We refer to the techniques used in mesh parameterization for this purpose.
While there are several techniques for curved surface flattening and transformation \cite{floater2005surface, sheffer2007mesh}, we adopted an isometric mapping (Isomap) ~\cite{tenenbaum2000global} based approach.
Isomap is a non-linear dimensionality reduction technique and several recent techniques \cite{caixeta2021robust, yousaf2020extended, acosta2016geodesic, zhou2004iso} build on this core idea to flatten triangularized mesh surfaces.
This approach particularly suits our purpose as the algorithm keeps the geodesic distance between the points of the writing unchanged.
Although Isomap is not the fastest algorithm for surface flattening, it provides robustness and applies to surfaces with sharp curves that occasionally appear on our unsupported writing surface.
\new

\begin{table*}[hbt]
\small
  \caption{\todo{Comparison with prior works on acoustic motion tracking.}}
    \vspace{-0.1in}
  \label{tab:related_work}
  \begin{threeparttable}
  \begin{tabular}{ccccccccc}
    \hline
    {\bf System}&{\bf Setup}&{\bf Tech.}&{\bf Audible}&{\bf 1D/3D Accu.}&{\bf Refresh rate}&{\bf Range}&{\bf Mic sep.}&{\bf Compt.}\tnote{a}\\
    \hline
    CAT~\cite{mao2016cat}  & Speaker-Phone & FMCW & N & 4mm/9mm & 25Hz & 7m & 90cm & No\\
    SoundTrak~\cite{zhang2017soundtrak} & Speaker-Watch & Phase & Y & --/13mm & 86Hz & 20cm & 4cm & No\\
    MilliSonic~\cite{wang2019millisonic} & Microphone-Phone & FMCW + phase & N & 0.7mm/2.6mm & 40Hz & 3m & 6-15cm & No\\
    {\bf \name} &  {\bf Alexa-speaker} & {\bf CFCW + phase} & {\bf N} & \textbf{0.07mm/1.4mm} & {\bf 333Hz} & {\bf 70cm} &  {\bf 3.6cm} & {\bf Yes}\\
    \hline
\end{tabular}
\begin{tablenotes}\footnotesize
\item [a] \small The column of 'Compt.' indicates if the solution is compatible with voice interfaces on low sampling rate devices like Amazon Alexa.
\end{tablenotes}
\end{threeparttable}
\end{table*}

\section{Implementation}
\begin{figure}[h]
\begin{center}
\includegraphics[width=1.5in]{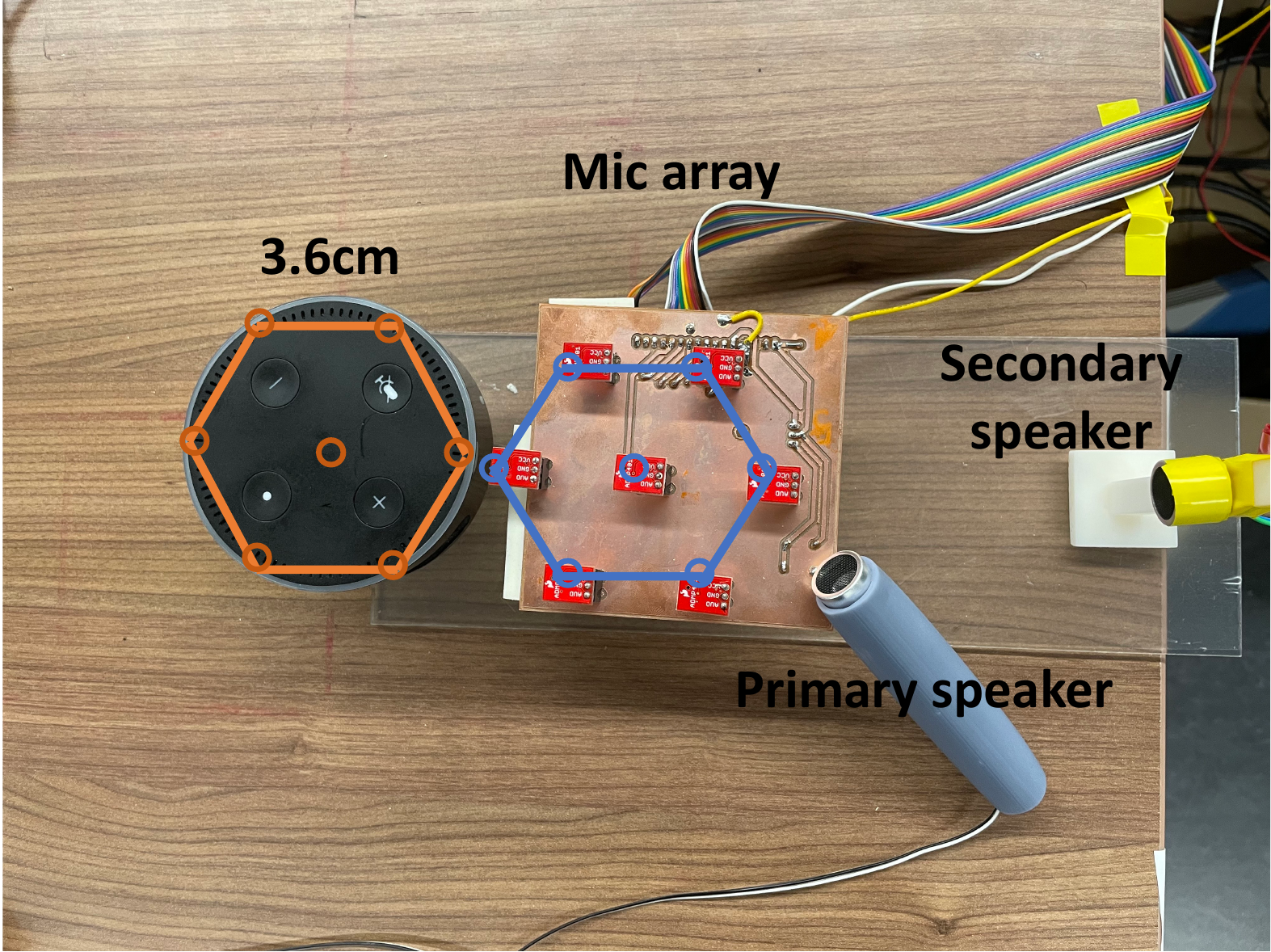}
\includegraphics[width=1.5in]{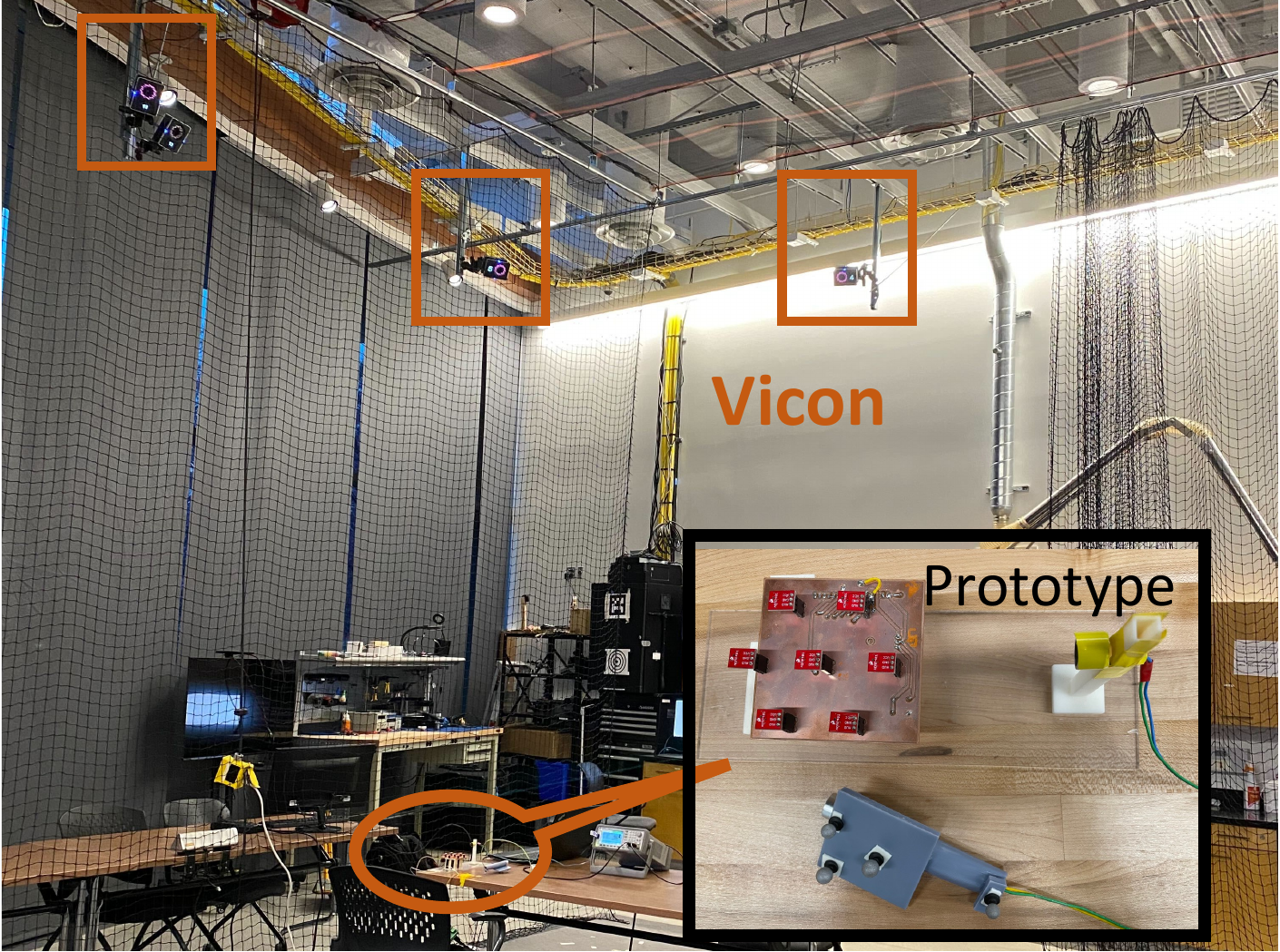}
\vspace{-0.1in}
\caption{(a) Hardware setup for system evaluation and (b) Vicon system for ground truth of 3D accuracy.}
\vspace{-0.1in}
\label{fig:hardware_new}
\end{center}
\end{figure}

To validate our implementation of \name, we build a hardware prototype and data processing pipeline. 
As shown in Figure~\ref{fig:hardware_new}(a), we use a 7-microphone array placed horizontally on the table and two low-cost(\$0.5) ultrasound speakers \cite{ultrasoundspeaker}.
The size of the microphone array is the same as Amazon Echo, with distances between microphones as 3.6cm.
We used omnidirectional ADMP401 MEMS microphones~\cite{devices2013admp401} sampled at 16kHz, 
\todoyang{similar to the components used in the Amazon device~\cite{alexaMic}}.
The microphones are sampled simultaneously using a multi-channel data acquisition system~\cite{Keysight}.
The secondary ultrasound speaker is placed at the same plane as the microphone array, 15cm away from the center.
\todo{The primary speaker is equipped with a pen-shape stylus or act as a transmitter to project signal in the air.}
The speakers are driven \todo{and synchronized} by a Keysight 33500B function generator.
The collected data is processed offline using Matlab scripts on a computer.
For 3D accuracy evaluation, we use Vicon~\cite{vicon} system as our ground truth, as shown in Figure~\ref{fig:hardware_new}(b).
\new

\section{Evaluation}
Table \ref{tab:related_work} summarizes the performance of \name and compares with existing techniques. 
\new


\subsection{Location tracking performance} 
\begin{figure}[h]
\begin{center}
    \includegraphics[width=1.5in]{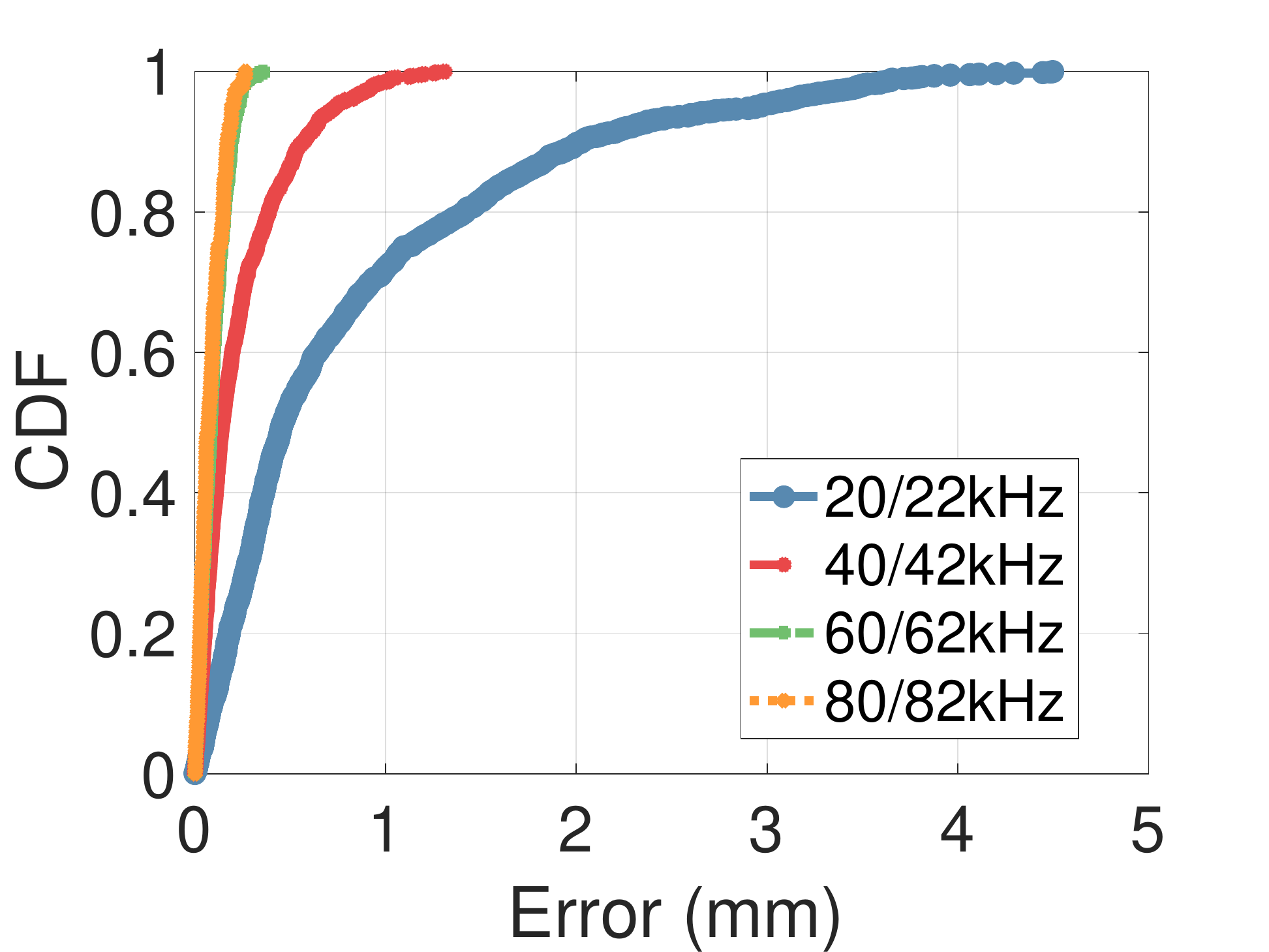}
    \includegraphics[width=1.5in]{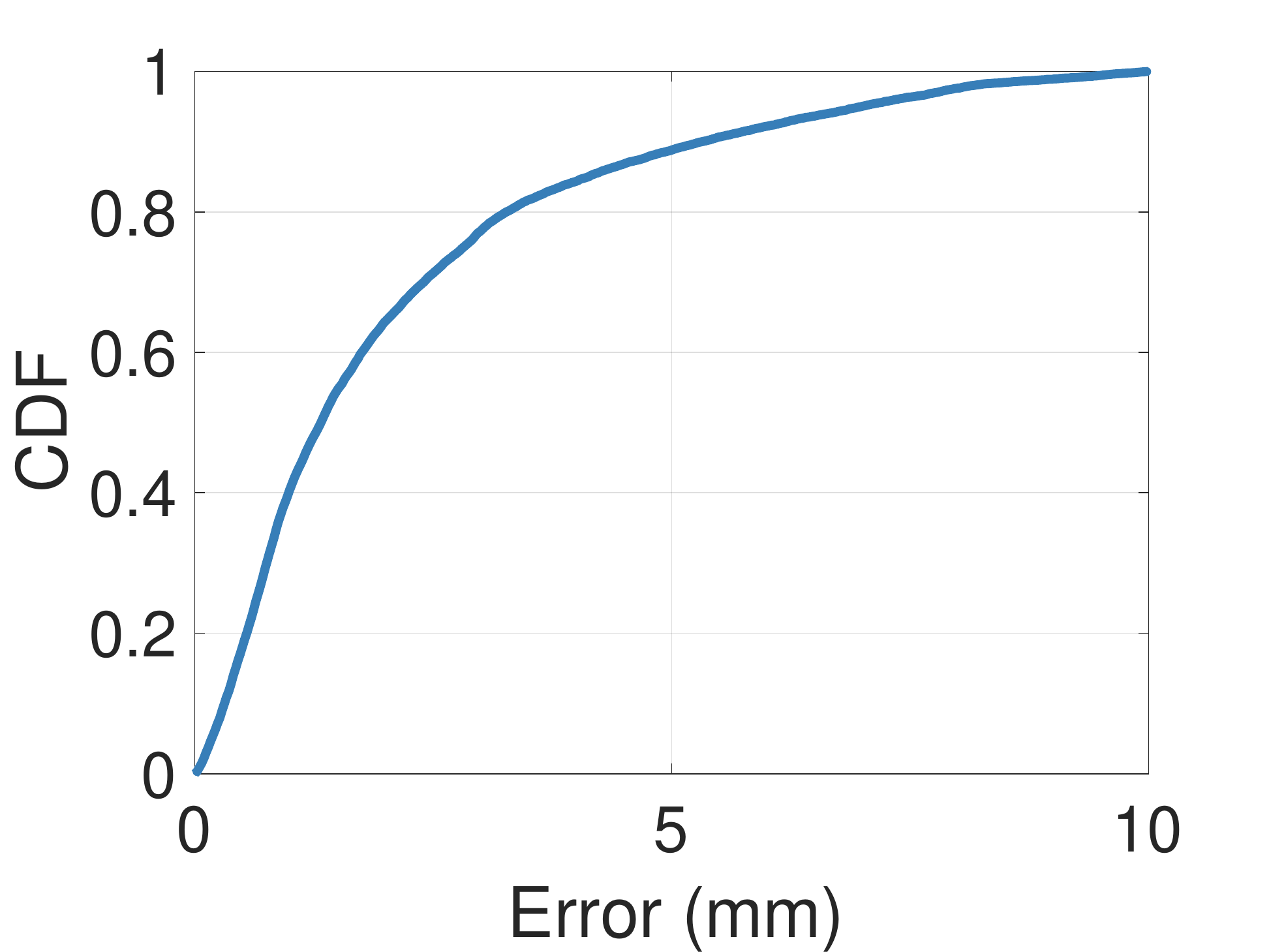}
    \vspace{-0.1in}
    \caption{The CDF of (a) ranging error and (b) 3D location tracking error of \name.}
    \vspace{-0.1in}
    \label{fig:accuracy_1d_3d}
\end{center}
\end{figure}

\textbf{Ranging accuracy.}
To get an accurate ground truth, we mount the primary speaker on a linear actuator~\cite{LinearActuator} that has a movement accuracy of $11\mu m$ to get precise ground truth.
The speaker moves from 10cm to 40cm. To evaluate the impact of frequency, for each frequency for the primary channel, we collect data 5 times.
We maintain a 7kHz separation between the primary and the secondary frequencies.
Figure~\ref{fig:accuracy_1d_3d}(a) shows the cumulative distribution function (CDF) plot of the ranging error with frequencies 20kHz, 40kHz, 60kHz, and 80kHz. 
\todo{The signals are transmitted by transmitters resonate at 20kHz~\cite{20kHzspeaker}, 40kHz~\cite{ultrasoundspeaker}, 60kHz~\cite{60kHzspeaker}, and 80kHz~\cite{80kHzspeaker}.}
Note that there is no non-linearity of the microphone for 20kHz signal, so we directly estimate the distance from the 20kHz frequency.
The frequency was hopped between the base frequency and 2kHz higher.
This plot shows that the accuracy increases with the higher frequencies, with median errors of 0.48mm, 0.16mm, 0.076mm, and 0.073mm.
The accuracy difference between 60kHz and 80kHz is small. 
\name's range can be increased by increasing the power of the primary speaker.
We evaluate the performance with different distances of the stylus from the voice assistant in Section~\ref{sec:distance}.
\new

\begin{figure}[hbt]
\begin{center}
\vspace{-0.15in}
\includegraphics[width=1.5in]{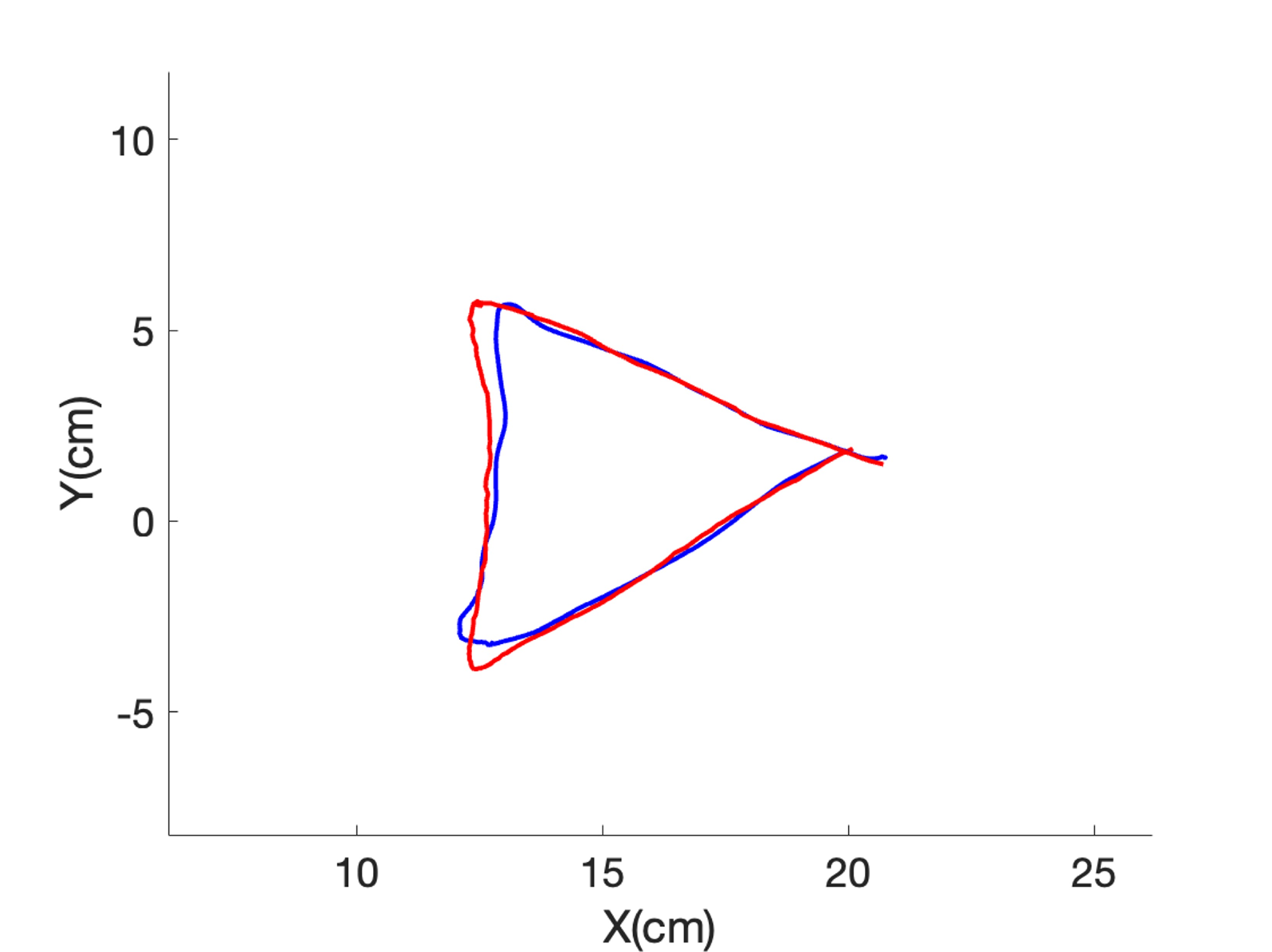}
\includegraphics[width=1.5in]{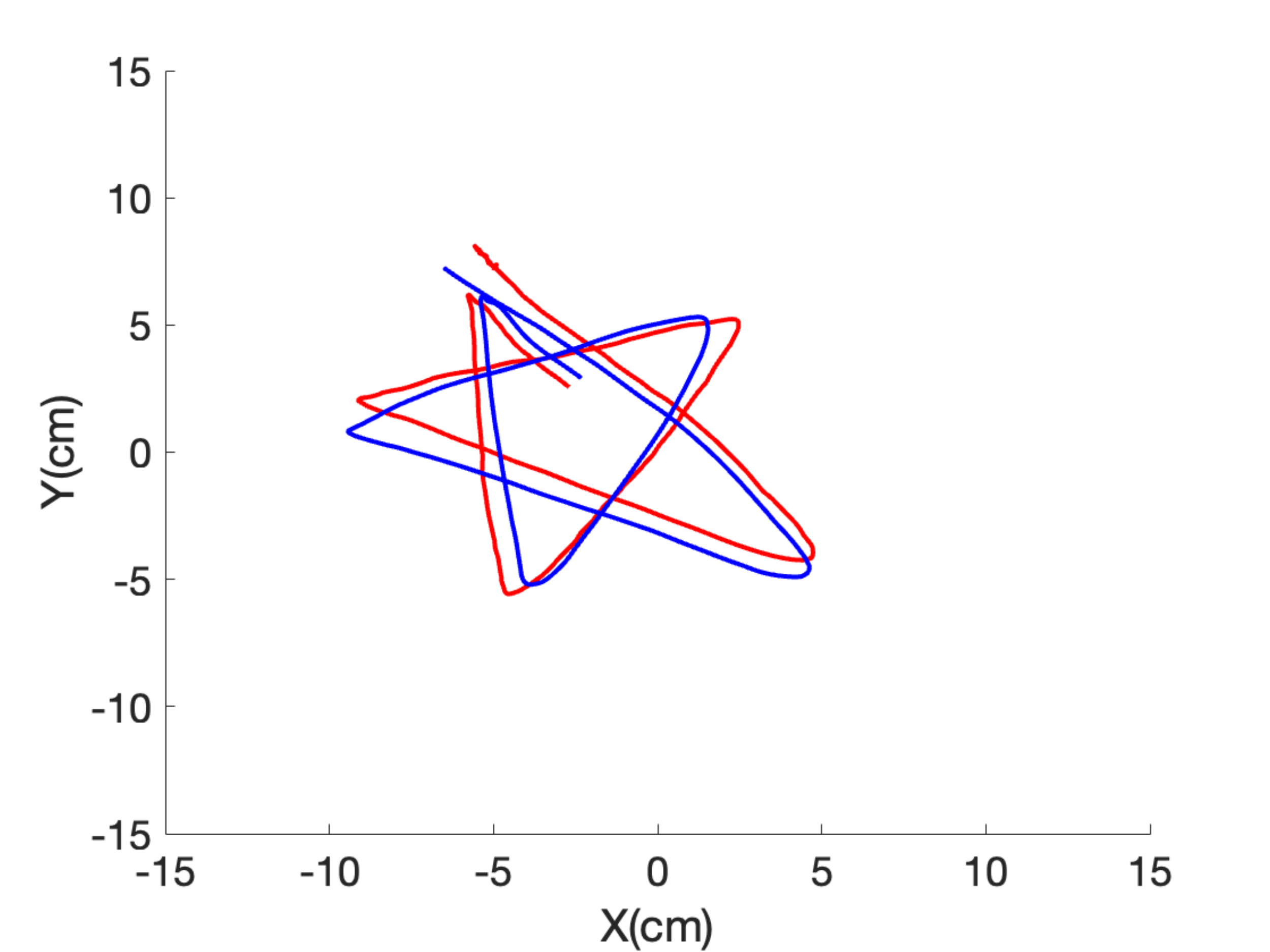}
\caption{3D tracking examples. The red and blue lines represent estimates from Vicon and \name respectively.}
\vspace{-0.2in}
\label{fig:shape_example}
\end{center}
\end{figure}

\textbf{3D location tracking accuracy.}
We measure the 3D localization and tracking accuracy with the distance estimations from 7 microphones in the array and the 3D localization algorithm described in Section~\ref{sec:loc_algo}.
To get the ground truth, we use the optoelectronic motion capture system of Vicon with 0.2mm tracking accuracy with a frame rate of 250Hz.
We also use the same frame rate so that the distance estimations can be matched.
The 3D region of detection is $0.5m\times0.5m\times0.4m$.
The frequency of the primary speaker is 40/42kHz.
Figure~\ref{fig:accuracy_1d_3d}(b) shows the CDF of 3D localization for \name.
The median error is 1.4mm with a 3.6cm distance between the microphones.
We evaluated \name with the similar microphone array configuration available in off-the-shelf voice assistants.
Figure~\ref{fig:shape_example} shows some qualitative examples of 3D tracking. 
Here the red lines indicate the ground truth trajectory captured by the Vicon system, and the blue line is the trajectory estimated with \name.

\subsection{Performance of voice recovery}
\begin{figure}[h]
\begin{center}
\vspace{-0.2in}
\includegraphics[width=1.5in]{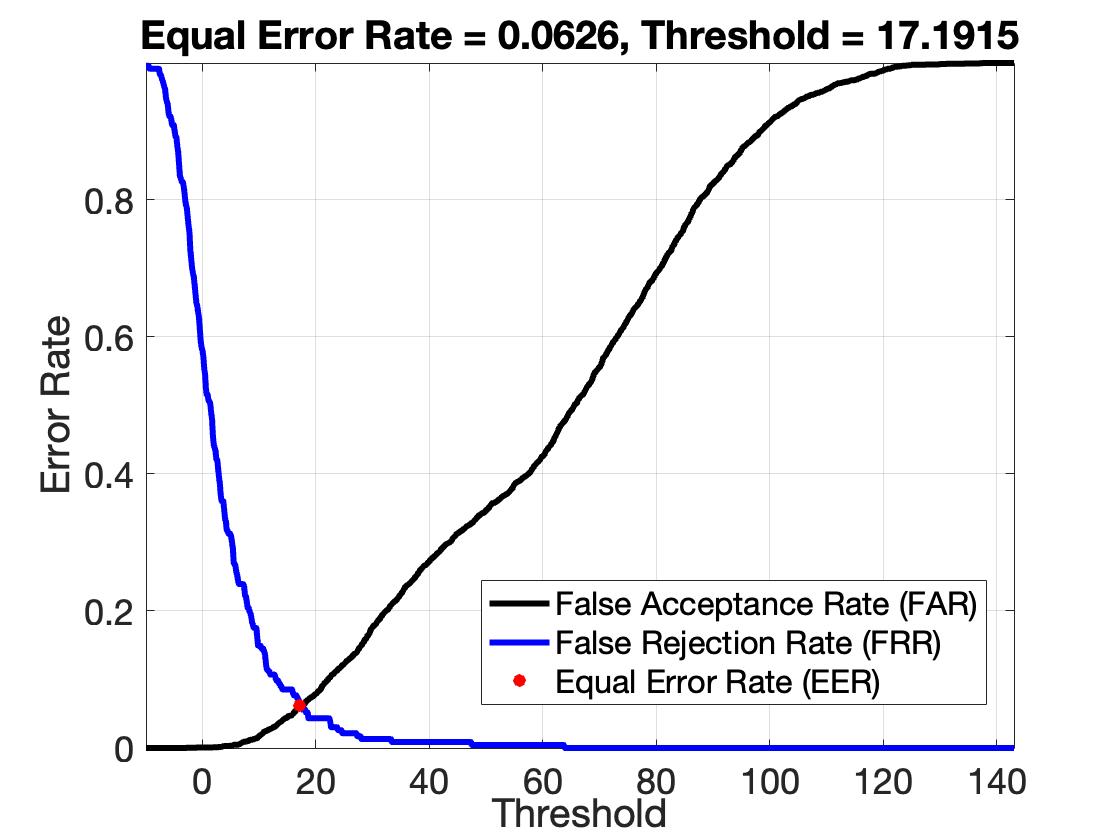}
\includegraphics[width=1.5in]{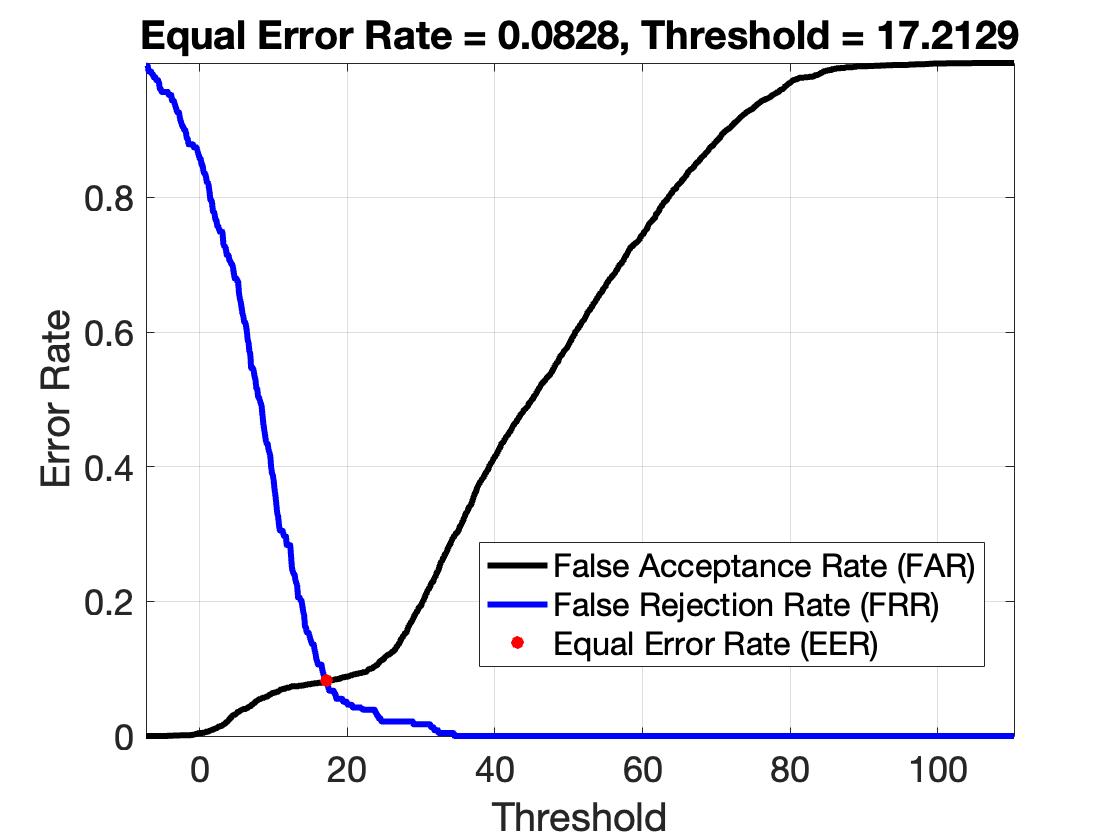}
\vspace{-0.1in}
\caption{Speaker authentication without (left) and with (right) the 7kHz handwriting tracking signal.}
\vspace{-0.2in}
\label{fig:speaker_authentication}
\end{center}
\end{figure}

\name is capable of simultaneously recovering voice and localization signals since \name only occupies 7kHz frequency, minimizing interference to the frequency band of voice. 
In this section, we evaluate the quality of the recovered voice in the presence of a ranging signal.
We apply an existing i-vector-based speech-independent speaker authentication~\cite{garcia2011analysis} and Google speech-to-text conversion~\cite{speechToText} techniques for the evaluation of voice recovery.
We first train the speaker authentication model with 9 males and 9 females in the Pitch Tracking Database from Graz University of Technology (PTDB-TUG)~\cite{database}, where each subject has 236 8-second voice pieces.
Then we enroll one male and one female using three voice samples. We use the rest 233 samples for testing.
To get the ground truth of the system, we first use the original voice pieces for testing.
After that, we play the samples while collecting the 7kHz handwriting tracking signal, and save them for testing.
We show the accuracy with 7kHz in Figure~\ref{fig:speaker_authentication}, the error rate only increases from $6.26\%$ to $8.28\%$, which indicates the speaker authentication system is robust despite simultaneous reception of handwriting tracking signal.
For speech-to-text conversion, we have one male and one female speaking 5 sentences each during the hand movement with and without \name. We feed the recorded signals to the speech-to-text converter without any change. The word recognition accuracies of converted text messages are both $96.0\%$. One example of the sentence is "Hi Alexa, I am using AirMarker to draw a star". The recognized text is "Hi Alexa, I am using hair marker to draw a star". 
\new

\subsection{Performance under human voice and environmental noise}
\begin{figure}[h]
\begin{center}
\includegraphics[width=1.5in]{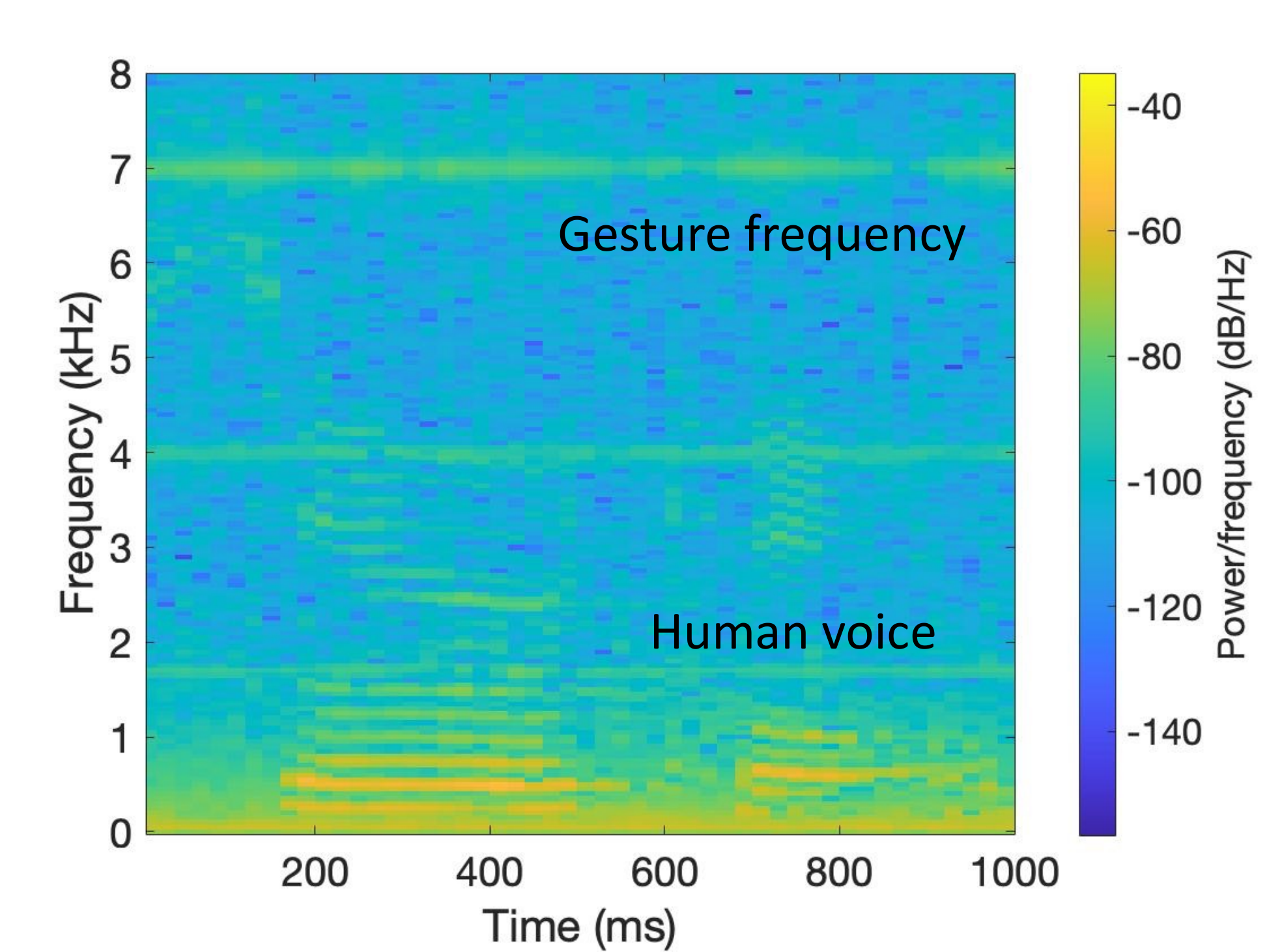}
\includegraphics[width=1.5in]{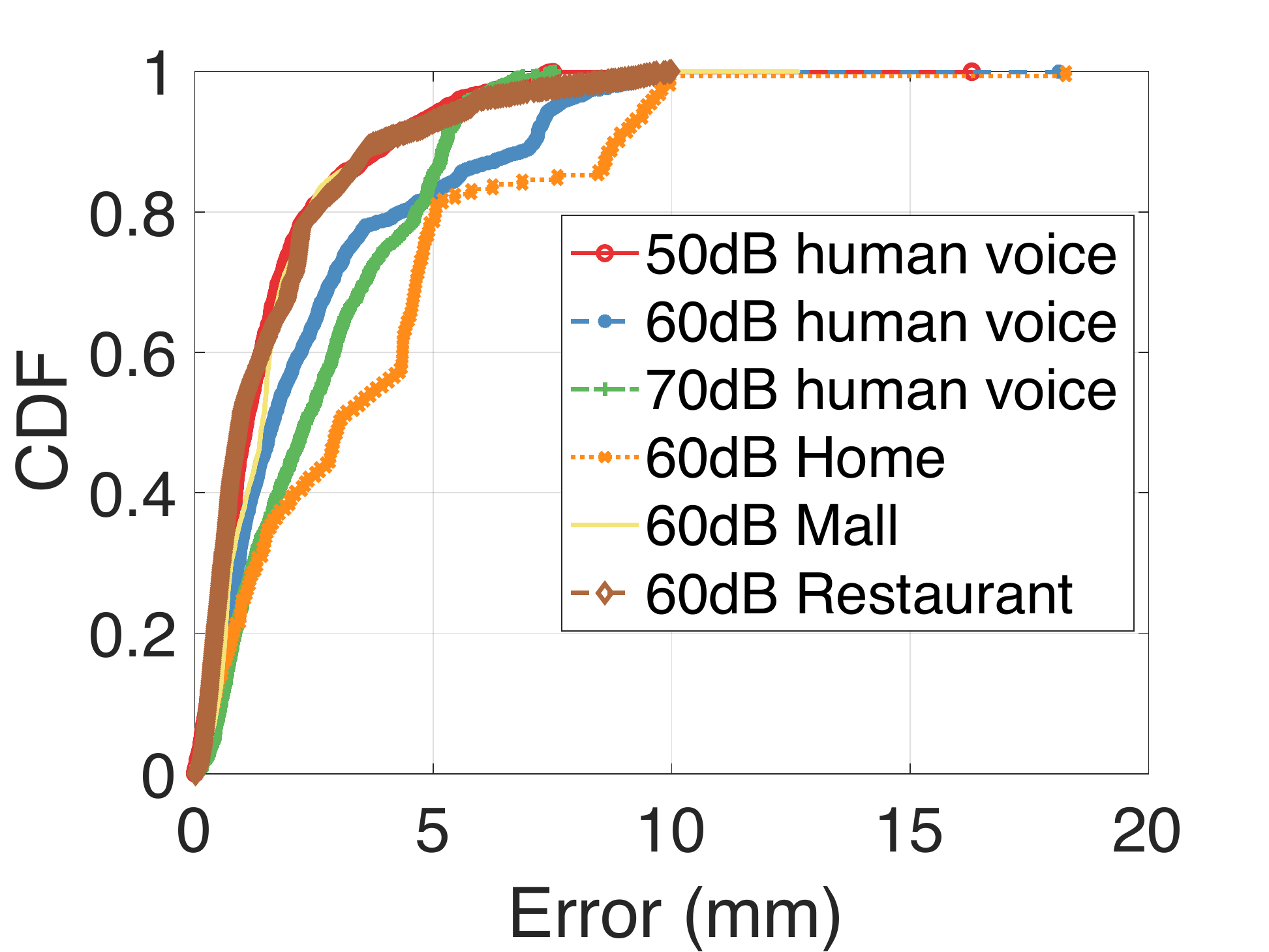}
\vspace{-0.1in}
\caption{Spectrogram of the recorded signal under human voice (left) and 3D results under different types of human voices and environmental noises (right).}
\vspace{-0.2in}
\label{fig:3d_with_voice_new}
\end{center}
\end{figure}

To evaluate the performance with human speaking while writing in the air, we perform the same process of evaluation as the 3D localization. The only difference is we play the noises at the same time.
We test the robustness under 50/60/70dB of human voices.
Note that the highest strength of human speaking in daily life is 50dB. 
Even when the sound level is 70dB, the median error of \name is 2.4mm, as shown in Figure~\ref{fig:3d_with_voice_new}(b), meaning the human voice does not affect its accuracy with a 16kHz sampling rate.
Figure~\ref{fig:3d_with_voice_new}(a) shows an example of the spectrogram of the recorded signal.
There is a clear separation between the frequencies of a human voice, environmental noise, and the 7kHz down-converted signal for tracking.
We also play three representative types of environmental noises in the home, mall, and restaurant.
\todoyang{The sound levels are 60dB.}
The median error of \name with noises in malls and restaurants are 1.4mm and 1.6mm, respectively.
The error under home noise is 3.4mm which is relatively high. 
The reason is that the frequency of running water has a component of 7kHz.

\subsection{Performance of handwriting recovery}
\begin{figure}[h]
\begin{center}
\vspace{-0.1in}
\includegraphics[width=1.6in]{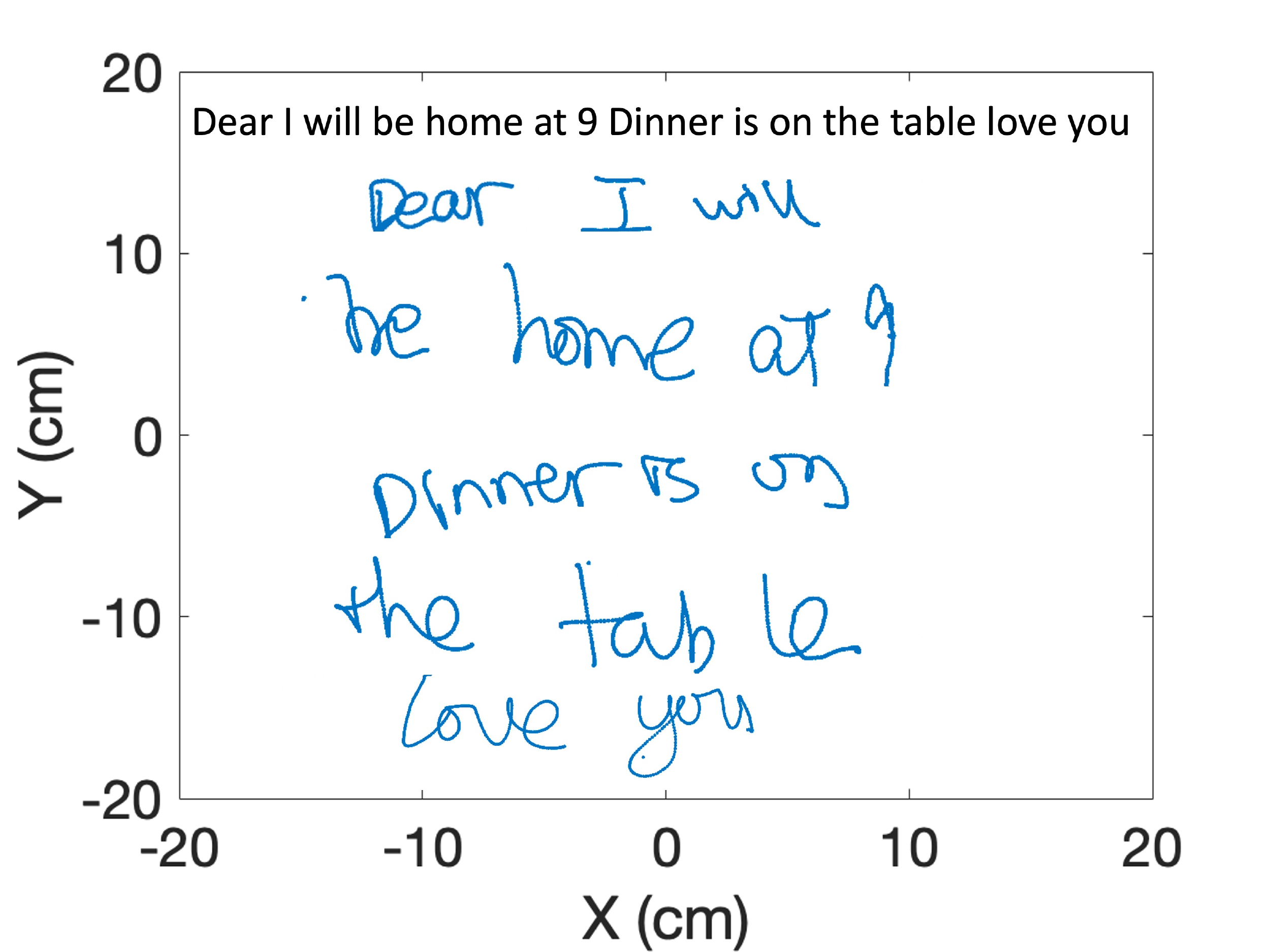}
\includegraphics[width=1.6in]{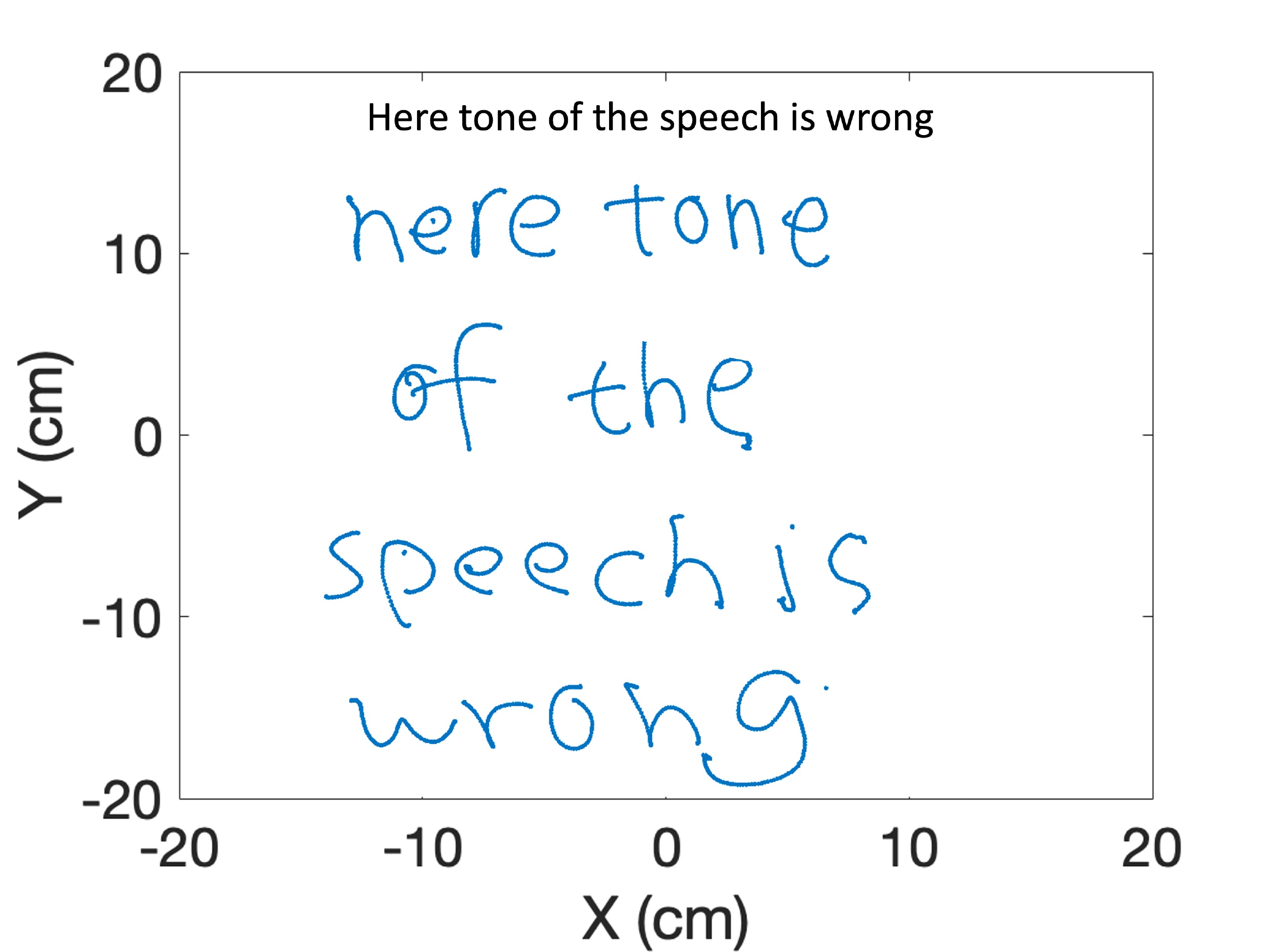}
\caption{Examples of paragraphs detection.}
\vspace{-0.2in}
\label{fig:paragraph_example}
\end{center}
\end{figure}

To test our performance of handwriting recovery, we write both paragraphs and single words in the air.
For the paragraph recovery, we write six representative paragraphs in the air and use Google Keep to convert images from text with an accuracy of $92.4\%$.
In Figure~\ref{fig:paragraph_example}, we show two representative examples.
To verify our accuracy of in-air writing-to-text conversion, we recruit ten participants (2 female and 8 male) to do a small-scale user study.
The users were asked to draw 5 words each.
Our objective is to compare the accuracy of writing to text conversion with handwriting on paper and in the air.
To enable a fair comparison, we attach the primary speaker to an apple pencil. 
After capturing both the on-paper and in-air writings, we feed them into Google Keep for writing-to-text conversion and compare the accuracy. 
We compare the accuracy of each word from the ten users, which are "alexa", "home", "okay", "sign", and "word".
The accuracy is the number of correctly detected characters over the number of characters.
We find the overall accuracy of \name and on-paper writing are $94.1\%$ and $96.6\%$, which means \name shows comparable accuracy with writing on paper.

\subsection{Performance of signature recovery}
\todoyang{To test the performance of signature recovery, we \todo{do a survey from both professional and general public.}
Our objective is to compare the similarity of the handwritten signatures and in-the-air recovered signatures. 
Not only the written text can be recognized, but the biometric characteristics in handwritten signature actions can be accurately captured, so that signature verification is possible.}
\todoyang{
For human study verification, we first recruit 20 participants (4 female and 16 male) to sign their signatures both on an iPad using an apple pencil, and in the air. 
To avoid the inconsistency between the two signatures, we attach the primary speaker to the apple pencil as the ground truth.
}
\new

{\bf (a) Professional signature verification:}
\todoyang{After capturing both on-tablet and in-air writings, we first ask a signature verification professional to give his opinion on if the 20 pairs of signatures are matched or not.
The professional treats 18 out of 20 pairs are matched, resulting in an accuracy of $90\%$.
The strategy of signature matching follows two rules: (1) the structure of the signatures should be the same, including the shapes of the whole signature and each character. (2) The angles of the pen touching and lifting the surface should follow the same trend.
}


\new

\begin{figure}[h]
\begin{center}
\includegraphics[width=3.3in]{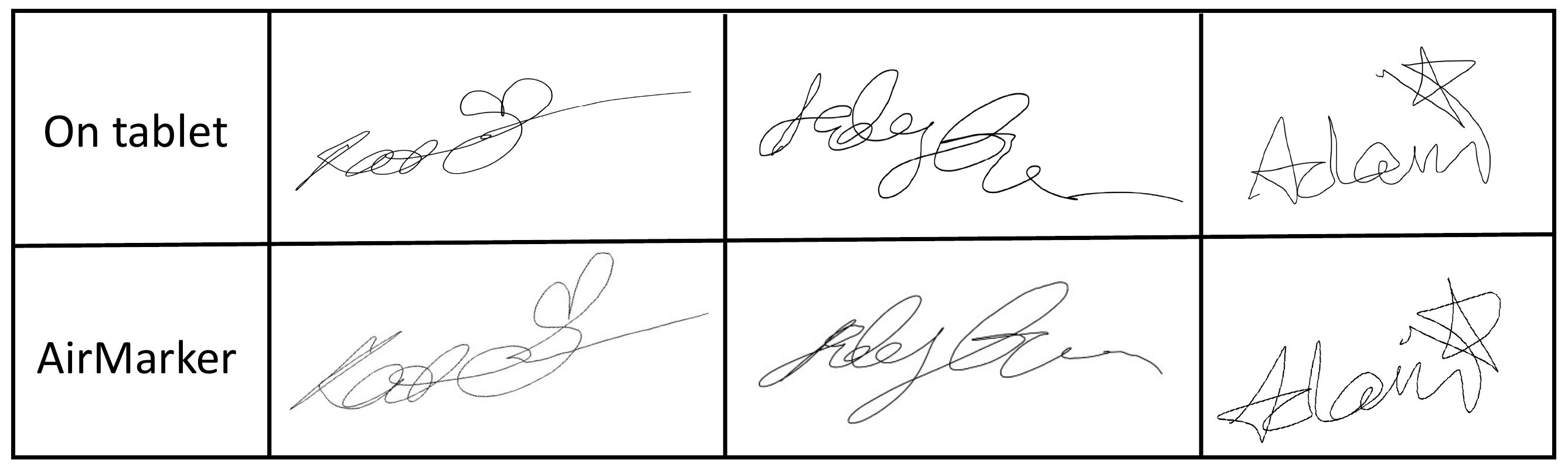}
\vspace{-0.1in}
\caption{\todoyang{Examples of signature pairs on iPad tablet (top) and recovered in-air signature (bottom)}.}
\vspace{-0.1in}
\label{fig:signature_example}
\end{center}
\end{figure}

{\bf (b) Opinion of general public:}
\todoyang{After that, we select five pairs of signatures and post a survey on Amazon Mechanical Turk.
30 people participated in the survey to share their opinions on if the in-air signatures are real or forged. 
Among the 150 ratings, $85.3\%$ of them are more than $50\%$ sure that the pair of signatures are from the same person. 
This result means \name not only can capture the recognizable written characters but also can capture precise biometric characteristics of handwritten signature actions, which can be used for signature verification.
Several examples of signatures are shown in Figure~\ref{fig:signature_example}.}


\begin{figure}[h]
\begin{center}
\includegraphics[width=1in]{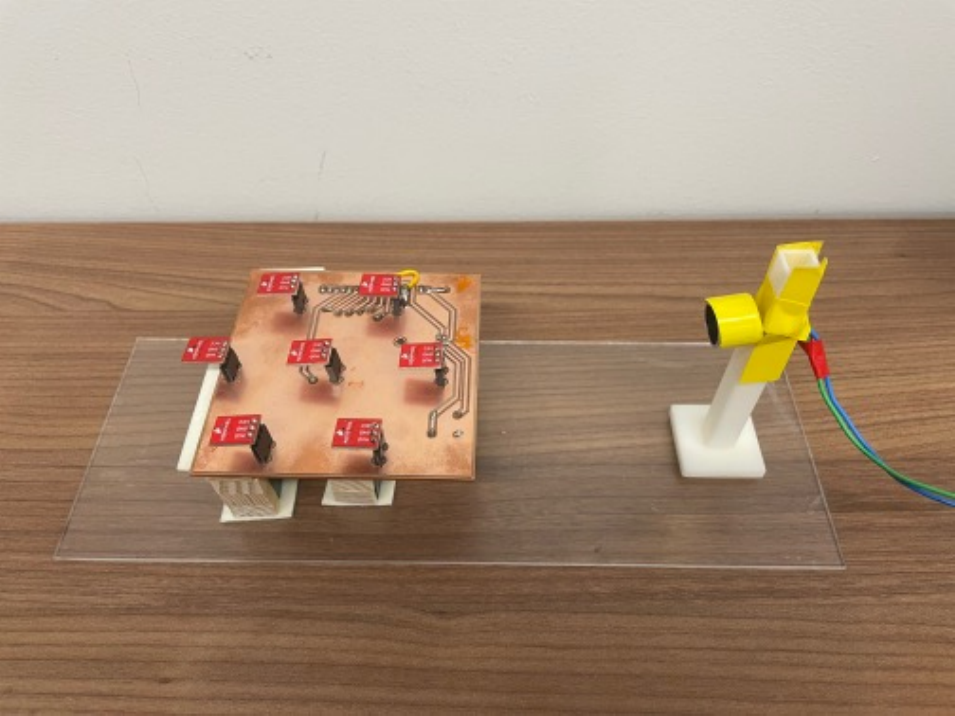}
\includegraphics[width=1in]{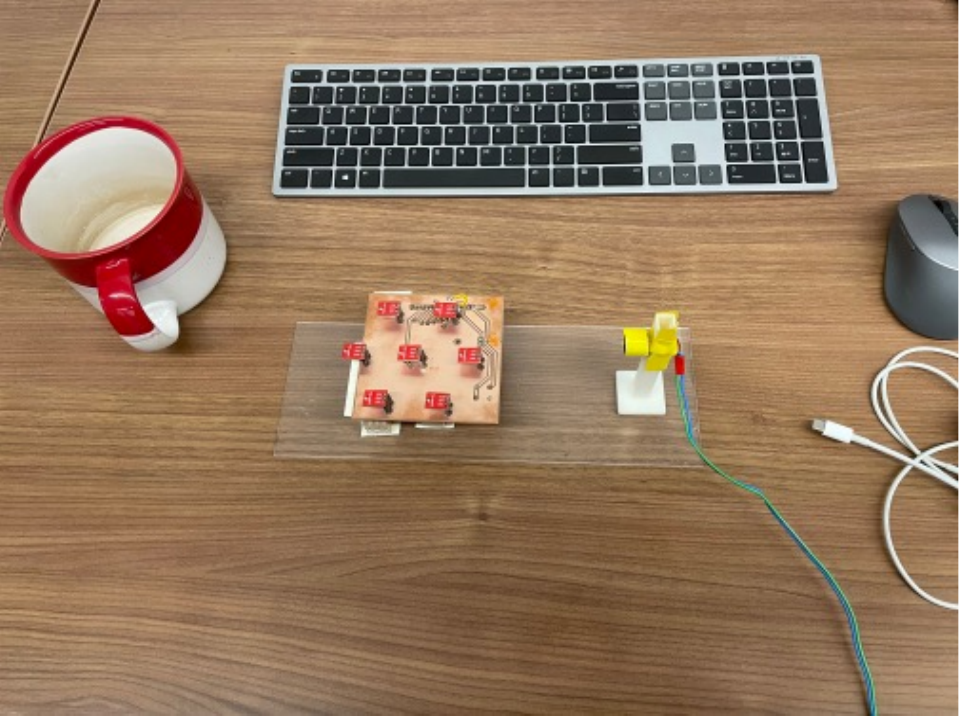}
\includegraphics[width=1in]{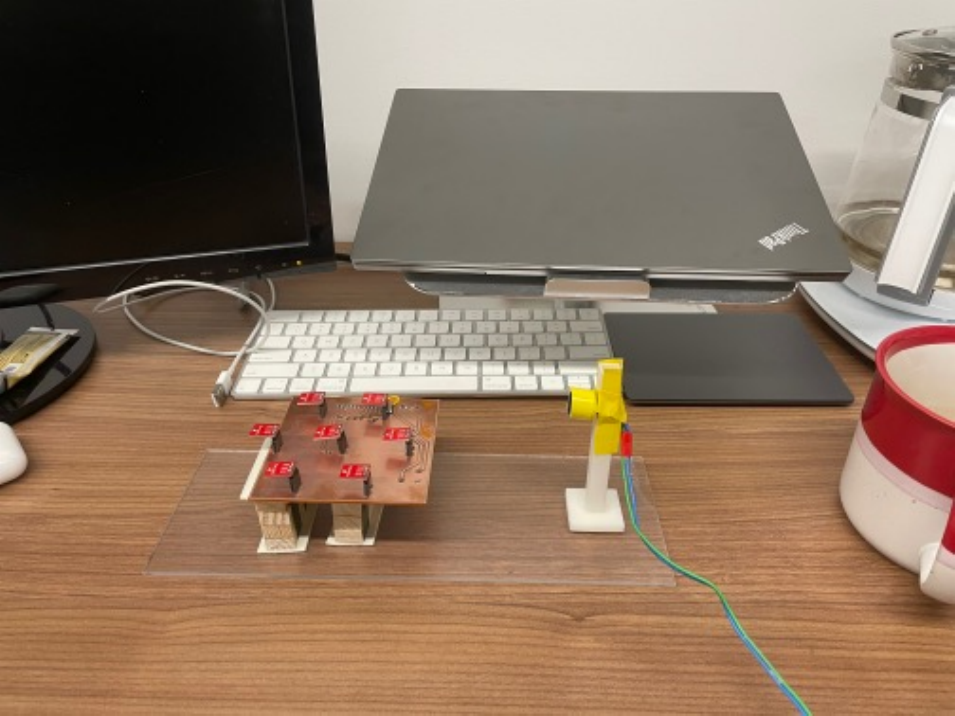}
\caption{Scenes of different multipath (a) wall, (b) clutters (3) both clutter and wall.}
\vspace{-0.1in}
\label{fig:scene_new}
\end{center}
\end{figure}

\subsection{Robustness to multipath}
\begin{figure}[hbt]
\begin{center}
\vspace{-0.1in}
\includegraphics[width=1.6in]{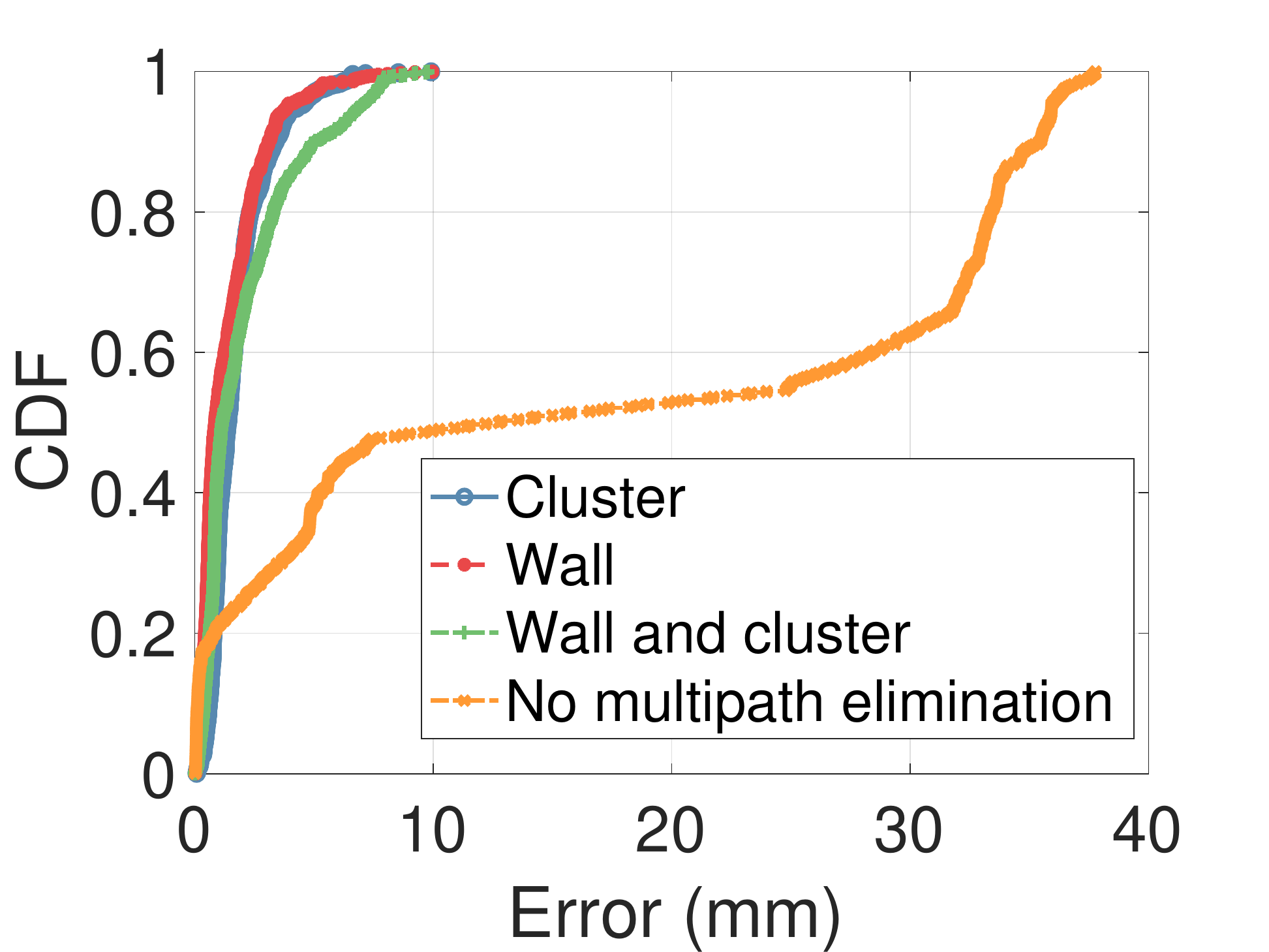}
\includegraphics[width=1.6in]{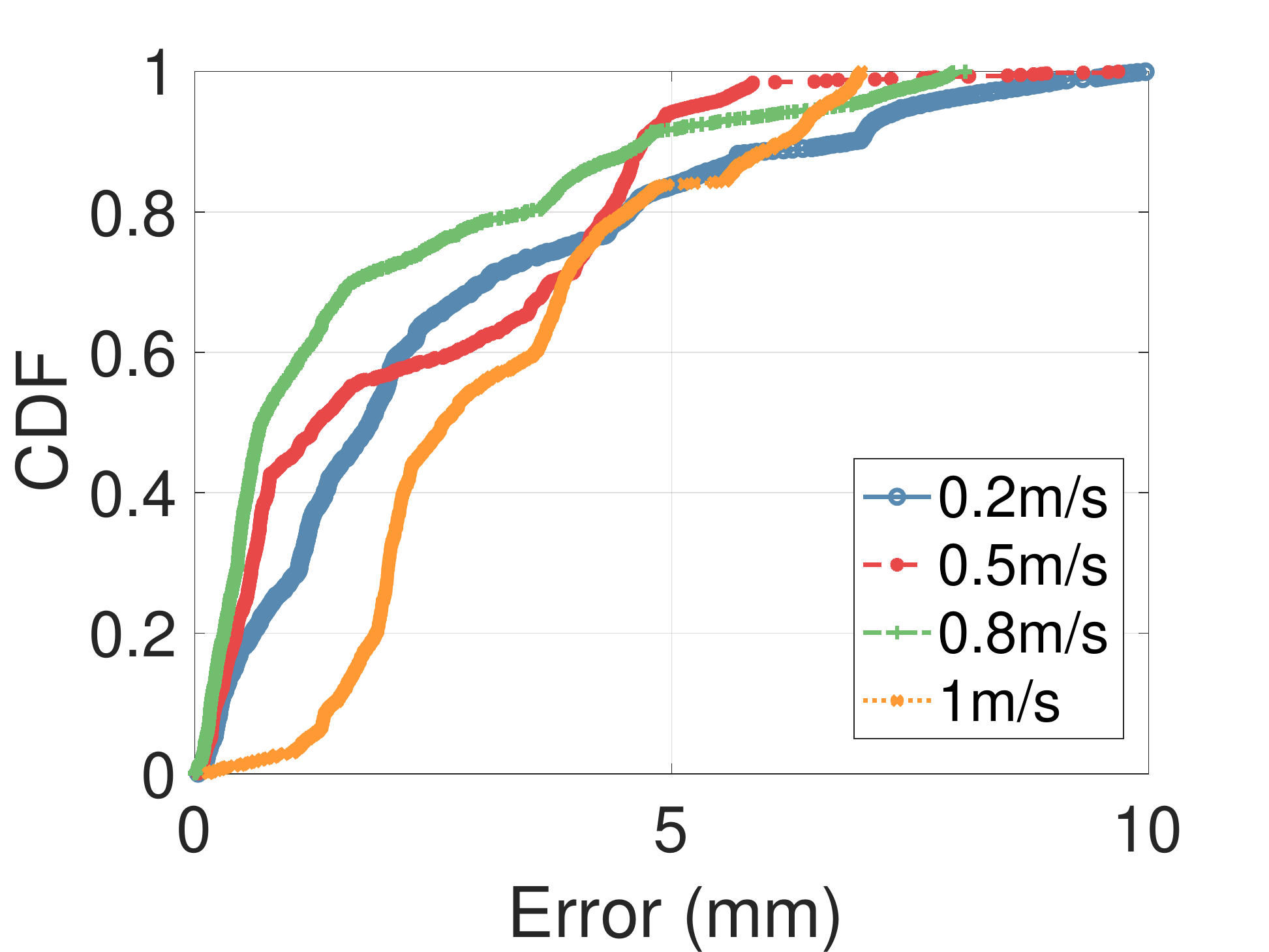}
\vspace{-0.1in}
\caption{(a)Comparison of the performance under multipath with/without CFCW for multipath elimination. (b)Performance of \name with different frequency resolutions under Doppler shift.}
\vspace{-0.1in}
\label{fig:multipath}
\end{center}
\end{figure}

We consider the common multipath when putting Alexa at different locations in a room, including a wall in behind, in clutter, and both in clutter and a wall in behind.
The scenes are shown in Figure~\ref{fig:scene_new} and results in Figure~\ref{fig:multipath}(a).
When there is multipath, the median errors are still within 1.4mm, which means \name is robust to multipath.
To test the effectiveness of the cross-frequency approach for multipath elimination, we do the same experiment with a wall behind, but only using one pair of frequencies. As shown in Figure~\ref{fig:multipath}(a), the median error increases to 10mm, meaning our multipath elimination approach is effective.

\subsection{Robustness to Doppler effect}
To test the effect of moving speed, we draw three shapes at speeds of 0.2m/s, 0.5m/s, 0.8m/s, and 1m/s.
The frequency of the primary speaker is 40/42kHz.
The maximum frequency shift caused by the Doppler effect is 126Hz.
As shown in Figure~\ref{fig:multipath}(b), the median errors are all within 1.9mm when the speed is within 0.8m/s. 
The error increases to 2.6mm when the speed is 1m/s.
This result shows our system is still reliable with a 1m/s moving speed, which is an extreme case of hand movement in the air.


\subsection{Impacts of external conditions}

\begin{figure}[h]
\begin{center}
\includegraphics[width=1.5in]{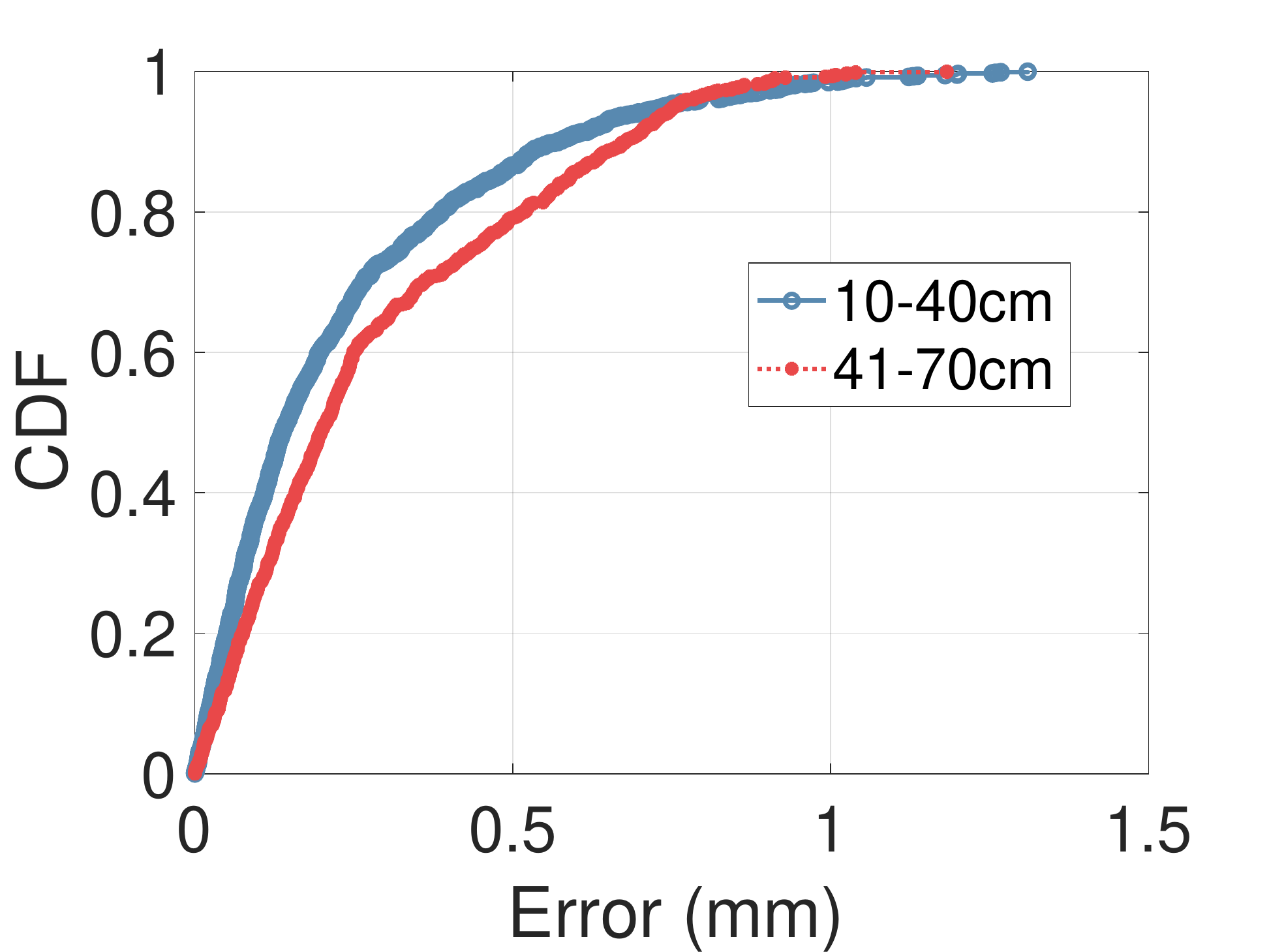}
\includegraphics[width=1.5in]{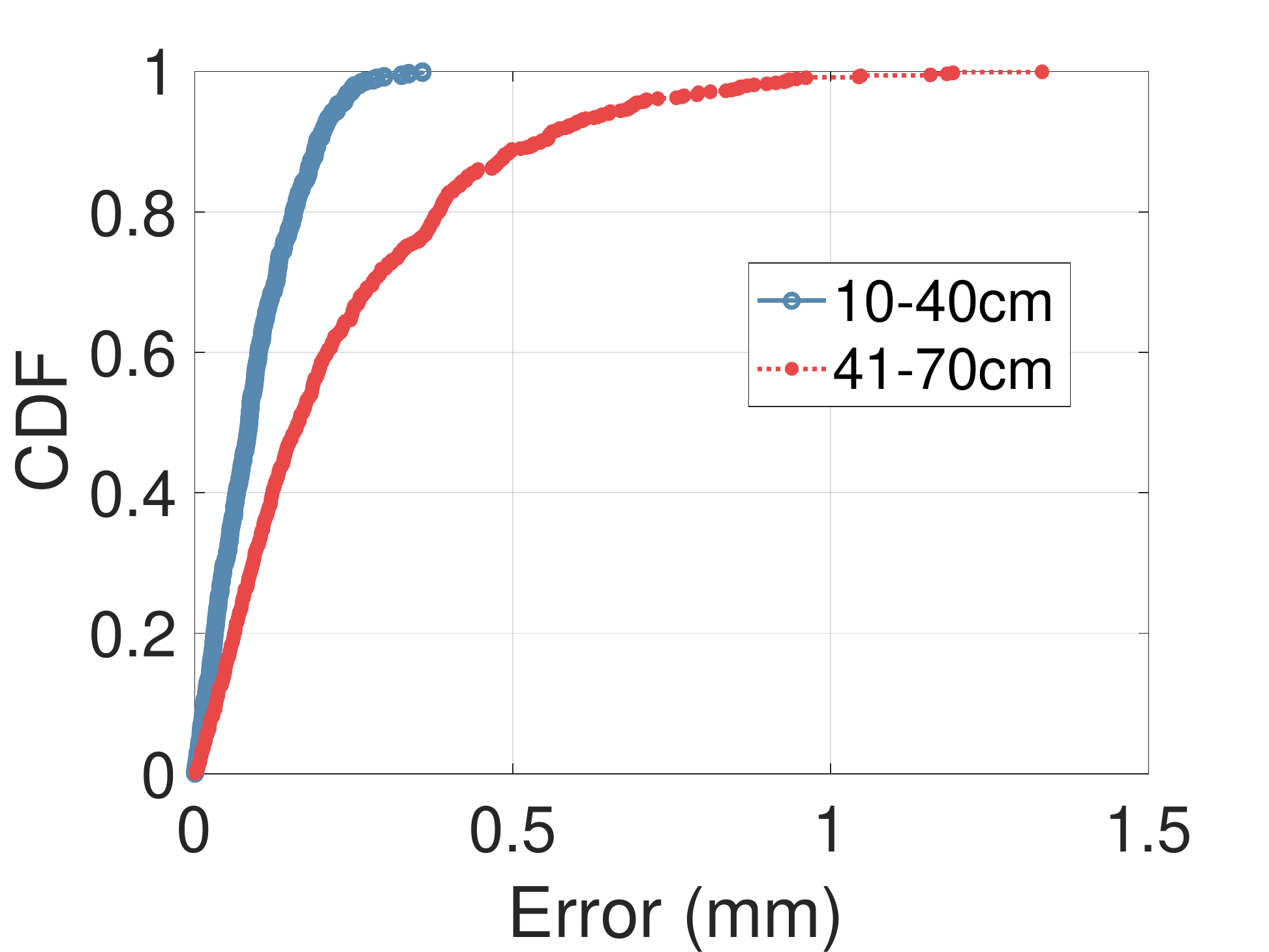}
\vspace{-0.1in}
\caption{1D results for 40/42kHz (left) and 60/62kHz (right) for different distances.}
\vspace{-0.1in}
\label{fig:distance_new}
\end{center}
\end{figure}

{\bf Distance to the microphone.}\label{sec:distance}
Figure~\ref{fig:distance_new} shows the CDF plot of 10-40cm and 41-70cm for 40/42kHz and 60/62kHz primary channel, respectively.
The plot shows that the system performs better in the 10-40cm range, with median errors of 0.14mm and 0.07mm for 40/42kHz and 60/62kHz frequencies.
It also confirms that higher frequency performs better following the theory.
When the distance increases to 41-70cm, the median errors are 0.23mm and 0.24mm.
The increase of error is because the SNR of the acoustic signal reduces with distance.
Moreover, when the distance increases, 60/62kHz frequency does not perform better than 40/42kHz.
The reason is the higher frequency signal attenuates faster than lower frequency signals.
Overall, the median error is still within 0.25mm in 41-70cm distance.

\begin{figure}[hbt]
\begin{center}
\includegraphics[width=1.5in]{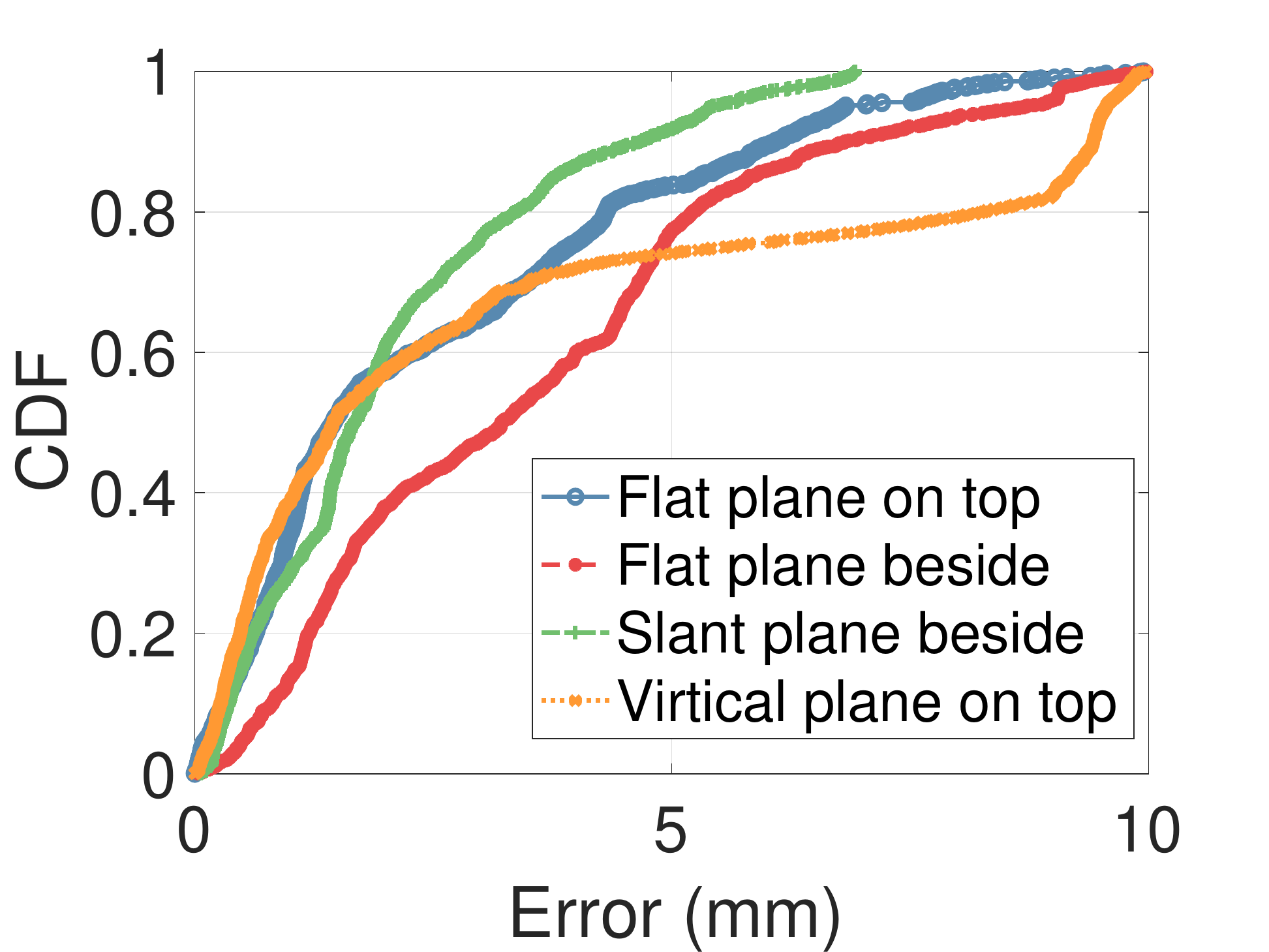}
\includegraphics[width=1.5in]{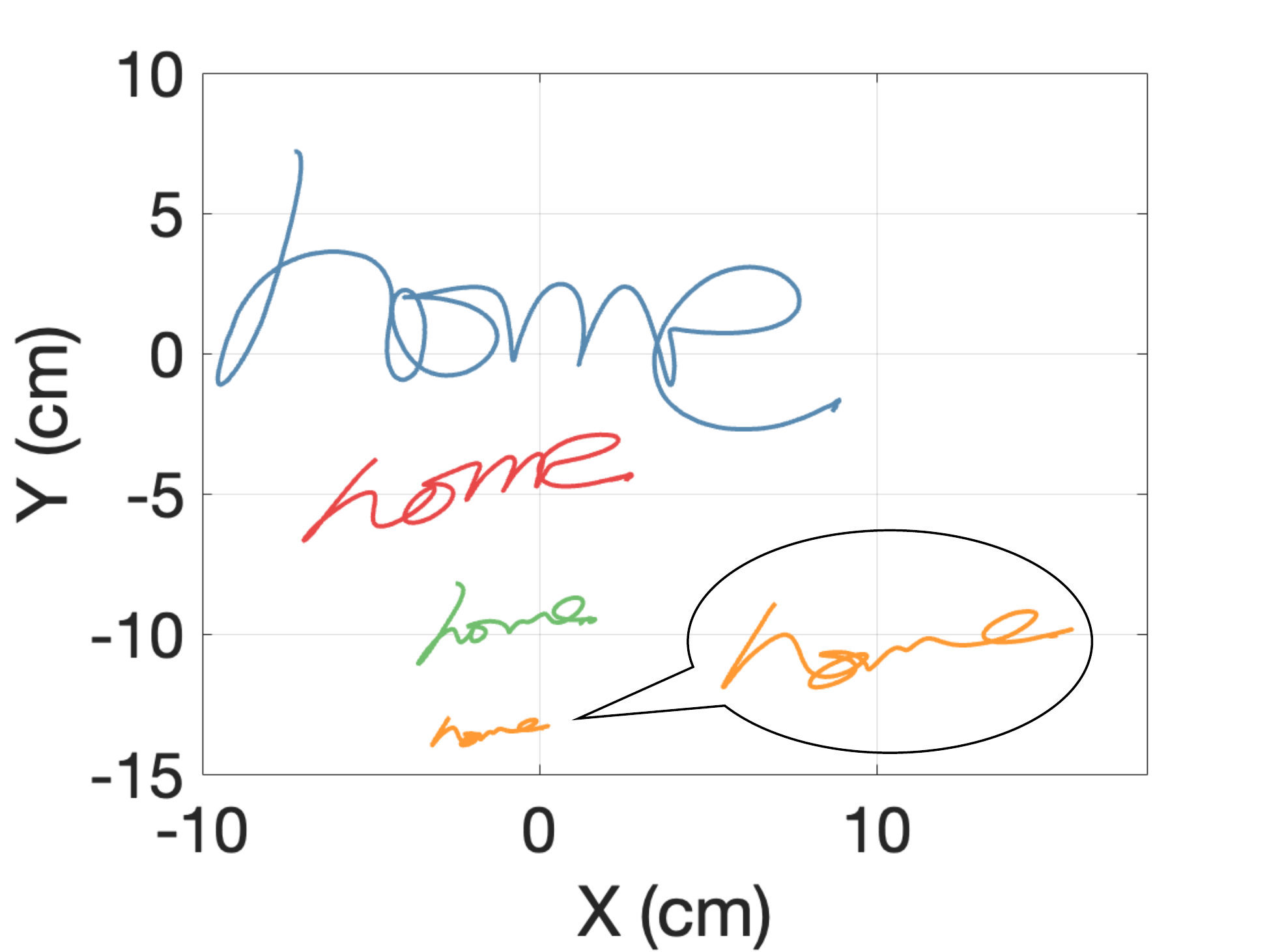}
\vspace{-0.1in}
\caption{Performance under different plane shifts (left) and size of written word (right).}
\vspace{-0.1in}
\label{fig:plane_size}
\end{center}
\end{figure}



{\bf Plane shift.}
We evaluate the performance when writing at different planes, including flat plane on top, flat plane besides, slant plane besides, and vertical plane on top.
As shown in Figure~\ref{fig:plane_size}(a), aside from writing on the flat plane beside the array, the median errors are all within 2mm. The median error of writing on the flat plane beside the array is 3.3mm.
The reason is when writing on the same plane with the microphone array, the pattern of distance changing is similar for all the microphones. Thus, same distance estimation error will cause a bigger location error.

{\bf Size of the word.}
To test the effectiveness of writing recovery with different sizes, we write the word "home" in widths from 3 to 20cm.
As shown in Figure~\ref{fig:plane_size}(b), we find that even in a width of 3cm (0.75cm per character), we still can recover the word clearly and can be recognized by Google Keep.
This result means \name is applicable for capturing regular sizes of writing for humans.

\section{Related Work}
The literature is rich in spatial analysis and localization. 
We sample below three topics closely related to this work.
\new

\textbf{Acoustic motion tracking.}
Acoustic signals are actively being explored by the research community for precise localization and tracking~\cite{wang2021mavl, cao2020earphonetrack, huang2014shake, yun2015turning}.
\todo{Existing studies detect minute body movements to monitor sleep apnea events~\cite{nandakumar2015contactless} and breathing for adults~\cite{xu2019breathlistener} and infants~\cite{wang2019contactless}.}
AAMouse~\cite{yun2015turning} applies Doppler shifts of acoustic signals to track hand movements in real-time.
CAT~\cite{mao2016cat} further enhances the tracking accuracy by analyzing both
FMCW and Doppler shift of acoustic signals. FMCW maps time difference to frequency shift, without the need for precise synchronization.
SoundTrak~\cite{zhang2017soundtrak} tracks accumulated phase shift for continuous tracking of a speaker in the air.
MilliSonic~\cite{wang2019millisonic} also applies the FMCW signal on hand tracking. To achieve sub-millimeter level accuracy, it tracks the phase shift of FMCW.
On the other hand, LLAP~\cite{wang2016device} and FingerIO~\cite{nandakumar2016fingerio} achieve device-free mm-level finger motion tracking. 
\todo{In our work, we explore the possibility of finer localization that does not interfere with voice interface on low sampling rate voice assistants.
The CFCW sonar technique used in \name enables household acoustic devices to achieve tens of micrometer-level motion tracking accuracy.}


\new

\textbf{RF and inertial sensor-based motion tracking.}
\todo{
We have seen rich literature on developing radio frequency techniques for localization and tracking, such as commercial
battery-free tags~\cite{jiang20193d,jin2018rf,jin2018wish,wang2013dude,wang2021locating,xiao2017one,yang2014tagoram} and custom RF backscatter~\cite{chang2018rf,chuo2017rf,kotaru2017localizing,ma20163d,nandakumar20183d,vasisht2018body,wang2019rfid}. 
However, they still cannot outperform acoustic motion tracking due to the tradeoff between latency, frame rate, and motion detection. 
Systems which can provide centimeter-level results either require a highly constrained environment~\cite{wang2013dude,yang2014tagoram} or have frame rates less than 1 Hz~\cite{ma2017minding,li2009multifrequency} since these systems need to step over hundreds of megahertz of bandwidth and take several seconds to compute a single location estimate.
More recently, novel systems have been proposed to significantly reduce the latency of localization systems to tens of milliseconds. However these systems either require using multiple antennas on the backscatter tags~\cite{nandakumar20183d} or using multiple RF transmitters by scanning over the different spectrum~\cite{luo20193d}.
Inertial sensor-based motion capture is based on miniature intertial sensors, sensor fusion algorithm, and biomechanical models~\cite{xsens,synertial,cao2021itracku}. 
Inertial sensors can only measure the motions, but not the absolute locations. 
}
\new
\textbf{Cross frequency sonar.}
We are not the first to exploit non-linearity to down-convert the frequency of the recorded signal.
The notion of exploiting non-linearity was originally studied in 1957 by Westervelt’s seminal theory~\cite{westervelt1957scattering, westervelt1951theory}. 
The closer use of the non-linearity to our System name is Backdoor~\cite{roy2017backdoor} which shows that ultrasound signals can be designed to become recordable by unmodified microphones.
\todo{More studies use non-linearity of microphone for side-channel attack~\cite{farrukh2021s3,ramesh2021acoustics,liu2020maghacker,zhang2017dolphinattack}, indoor localization~\cite{lin2019rebooting}, and communication~\cite{bai2020batcomm}.}

\vspace{0.05in}
\section{Conclusion}
This paper develops an acoustic motion tracking interface for voice assistants. \name proposes a high-resolution CFCW-based distance and tracking estimation algorithm leveraging the nonlinearity of the microphone. 
\todo{Moreover, \name is the first motion tracking work that can co-exist with voice interface on low sampling rate voice assistants.}
Evaluations of the prototype show a 1-D ranging error of 73 micro-meter and below 1.4 millimeters of median error in 3D trajectory tracking. This paper presents multiple benchmarks, pilot user studies and evaluation for some specific applications. The design of \name enables an interface on voice assistants for capturing handwritten notes, drawings, and signatures and can extend to a non-verbal mode of human-machine interaction.
\bibliographystyle{acm}
\bibliography{reference,reference_owlet, referenceIrtaza}

\end{document}